\documentclass[useAMS,usenatbib]{mnras}

\usepackage{graphicx}
\usepackage{amssymb}
\usepackage{array}

\usepackage{ulem}

\newcommand{\kms}{km\,s$^{-1}$}
\newcommand{\Te}{T$_{\rm e}$}
\newcommand{\HI}{\textrm{H}\,\textsc{i}}
\newcommand{\HII}{\textrm{H}\,\textsc{ii}}
\newcommand{\Halpha}{\textrm{H}\,$\alpha$}
\newcommand{\OH}{\textrm{O/H}}
\newcommand{\FeH}{\textrm{[Fe/H]}}
\newcommand{\Msun}{$\textrm{M}_{\sun}$}
\newcommand{\Msunyr}{\Msun{}\,yr$^{-1}$}

\newcommand{\Vh}{V_{\rm hel}}

\graphicspath{ {./figs/} }

\newcommand{\Mtrgb}{$\textit{F814W} = 26.448\pm0.040$\,mag}         
\newcommand{\Ctrgb}{$(\textit{F606W}-\textit{F814W})_{\rm TRGB} = 0.977\pm0.007$\,mag}    
\newcommand{\modulusLong}{$(m-M)_0 = 30.527\pm0.070$\,mag}          
\newcommand{\modulus}{$30.53\pm0.07$\,mag}      
\newcommand{\distance}{$12.75\pm0.41$\,Mpc}     

\title[Unusual void galaxy DDO\,68]
{Unusual void galaxy DDO\,68: implications of the \textit{HST}-resolved photometry}

\author[
D.\ I.\ Makarov,
L.\ N.\ Makarova,
S.\ A.\ Pustilnik
\&
S.\ B.\ Borisov
]
{D.\ I.\ Makarov,$^1$\thanks{E-mail: dim@sao.ru (DIM)}
L.\ N.\ Makarova,$^1$\thanks{E-mail: lidia@sao.ru (LNM)}
S.\ A.\ Pustilnik$^1$\thanks{E-mail: sap@sao.ru (SAP)}
and
S.\ B.\ Borisov$^{2,3}$
\\
$^1$ Special Astrophysical Observatory of RAS, Nizhnij Arkhyz, Karachai-Cherkessian Republic, 369167, Russia\\
$^2$ Sternberg Astronomical Institute, Moscow State University, Moscow, 119991, Russia\\
$^3$ Department of Physics, Moscow State University, 1, Leninskie Gory, Moscow, 119991, Russia
}

\begin{document}

\label{firstpage}

\pagerange{\pageref{firstpage}--\pageref{lastpage}}

\pubyear{2016}

\maketitle

\begin{abstract}

DDO\,68 (UGC\,5340) is an unusual dwarf galaxy with extremely low gas
metallicity [$12+\log(\OH) = 7.14$] residing in the nearby Lynx--Cancer void.
Despite its apparent isolation, it shows both optical and \HI{} morphological
evidence for strong tidal disturbance.
Here, we study the resolved stellar populations of DDO\,68
using deep images from the \textit{HST} archive.
We determined a distance of \distance{} using the tip of the red giant branch (TRGB).
The star formation history reconstruction reveals that about 60 per cent of stars
formed during the initial period of star formation, about 12--14\,Gyr ago.
During the next 10\,Gyr, DDO\,68 was in the quenched state, with
only slight traces of star formation.
The onset of the most recent burst of star formation occurred about
300\,Myr ago.
We find that young populations with ages of several million to
a few hundred million years are widely spread across various parts of DDO\,68,
indicating an intense star formation episode with a high mean rate of
0.15\,\Msunyr.
A major fraction of the visible stars in the whole system ($\sim80$\,per cent)
have  low metallicities: $Z = Z_{\sun}/50\textrm{--}Z_{\sun}/20$.
The properties of the northern periphery of DDO\,68 can be explained by
an ongoing burst of star formation induced by the minor merger of
a small, gas-rich, extremely metal-poor galaxy with a more typical dwarf galaxy.
The current TRGB-based distance of DDO\,68 implies a total negative
peculiar velocity of $\approx500$\,\kms{}.

\end{abstract}


\section{Introduction}
\label{sec:intro}

\begin{figure*}
\centerline{
\includegraphics[width=0.8\textwidth,clip]{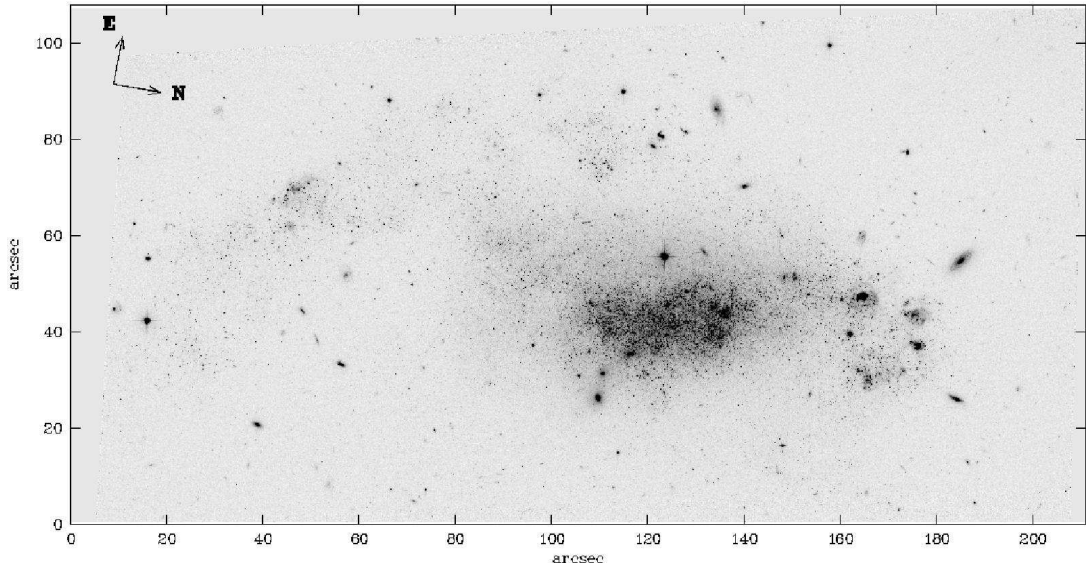}
}
\caption{
The \textit{HST}/ACS image of DDO\,68 in the \textit{F606W} filter. 
Only WFC1 detector field is shown.
}
\label{fig:ima}
\end{figure*}

The unusual dwarf irregular galaxy DDO\,68 (UGC\,5340) is located
in the nearby Lynx--Cancer void \citep{PT2011}.
Recent distance estimations place DDO\,68 between 12 and 13\,Mpc
\citep{DDO68C,TGL2014}.
The galaxy has a large negative peculiar velocity of about $-500$\,\kms{}.
This is $\sim200$\,\kms{} larger than the prediction of the velocity-field model
in this region by \citet{TSK2008}.

Despite its apparent isolation, DDO\,68 shows a peculiar morphology:
several prominent star-forming regions on the periphery,
mainly in the `northern ring' and the `southern tail' \citep{PKP2005}.
The \HI{} study with the Giant Meterwave Radio Telescope \citep[GMRT;][]{ECP2008},
as well as less sensitive observations with the
Westerbork Synthesis Radio Telescope \citep{SI2002},
show the complex gas structure and the velocity field consisting of two arms,
winding asymmetrically around the main bright part of the optical body.
\citet{ECP2008} suggest that DDO\,68 is likely the result of a
merging of two gas-rich dwarfs. The recent very deep \HI{} study with
the Very Large Array and the Green Bank Telescope revealed a faint interacting
companion DDO\,68C at a projected distance of 42\,kpc from DDO\,68.
The optical counterpart of the companion is hidden in the halo of a bright star,
preventing direct stellar mass measurement.
It was found that its gas mass is smaller than that of DDO\,68 by a factor of $\sim35$ \citep{DDO68C}.
\citet{Annibali+2016} concluded that the disturbed morphology of DDO\,68 and its nearby tidal features 
cannot be explained by the influence of DDO\,68C.

DDO\,68 has an extremely low gas-phase oxygen abundance of
$12+\log(\OH{})=7.14$--7.20.
The current \OH{}-value is based on direct measurements by the classic
\Te{}-method of the several brightest \HII{} regions in the `northern ring'
\citep{PKP2005,IT2007,Berg+2012,ITG2012}.
Along with its void `neighbour'
SDSS\,J0926+3343 with $12+\log(\OH{})=7.12$ \citep{PTK2010}
and the recently discovered very faint neighbour of the Local Group, Leo\,P,
with $12+\log(\OH{})=7.17$ \citep{Skillman+2013}, DDO\,68 is considered as
one of the most metal-poor galaxies in the Local
Volume and its surroundings.
A prototype extremely low metallicity blue compact galaxy I\,Zw\,18
has $12+\log(\OH{})=7.16$--7.19 \citep[e.g.,][and references therein]{Skillman+1993,Izotov+1998}.
It is somewhat farther, with distance estimates from $D \sim 15$ to 19\,Mpc \citep{IT2004,Contreras+2011}.
A couple of new, nearby, extremely metal-deficient dwarfs, UGC\,772 and SDSS\,J1056+3608,
with $12+\log(\OH{})\sim7.16$ were presented by \citet{ITG2012}.
The galaxy SBS\,0335$-$052W has the lowest known metallicity $12+\log(\OH{}) \sim 6.9$ \citep{Izotov+2009}.
However, it is a significantly more distant object, with a distance of about 54\,Mpc.

\citet{PT2011} emphasize the high concentration of very-low-metallicity dwarfs in the Lynx--Cancer void.
Recently, \citet{Hirschauer+2016} discovered the record-low metallicity for the
dwarf galaxy SDSS\,J094332.35+332657.6  [$12+\log(\OH{})=7.02\pm0.03$]
identified with the \HI{}-source AGC\,198691,
which was found within the framework of the blind \HI{} Arecibo Legacy Fast ALFA Survey \citep{ALFALFA40}.
One more extremely metal-poor dwarf galaxy J0706+3020 with ($12+\log(\OH{})=7.03$) 
was found near the centre of the void \citep{CPE2016,PPK2016}.
These findings reinforce the conclusion of \citet{PT2011} on the high concentration of very metal-poor galaxies in voids.
At a kinematic distance of $\sim10.8$\,Mpc,
AGC\,198691 appears to be a neighbour of both extremely low-metallicity galaxies, SDSS\,J0926+3343 and DDO\,68.

Using the \textit{Hubble Space Telescope} (\textit{HST}) archive data,
\citet{TGL2014} distinguish two very different spatial
components of the resolved stellar populations in DDO\,68. 
The population with medium metallicity ($Z_{\sun}/5$) is concentrated
in the central high density part of the galaxy. The low-metallicity
component ($Z \la Z_{\sun}/20$) winds around the main body from
the `northern ring' to the `southern tail'.
The authors found that the secondary component has only a small fraction
of old stars, as opposed to the case of the main body.

In the course of revising our paper, \citet{SAC+2016} published
their article on the star formation history (SFH) reconstruction in DDO\,68.
They conclude that almost 80 per cent of the total stellar mass ($M_*\simeq1.3\times10^8$\,\Msun{})
comprises of old (age $>1$\,Gyr) stars.
This result allows the authors to reject the hypothesis
that DDO\,68 is a young system experiencing its first burst of star formation.

In this work, we concentrate on the general study of stellar populations
of DDO\,68 using the deep images from the \textit{HST} archive.
We examine the distribution of stars of different ages and metallicities throughout the galaxy.
We reconstruct the SFH in different regions of
DDO\,68 using the Padova theoretical isochrones of stellar evolution
\citep{Padova2000}.
We determine the distance modulus with the improved tip of the red
giant branch (TRGB) method \citep{TRGB1} and calibration \citep{TRGB2}.

In a complementary paper \citep{DDO68LBV}, 
we analyse the variability of the known luminous blue variable in DDO\,68 \citep{DDO68-LBV,IT2009}.
Using the \textit{HST} \Halpha{} images,
high-resolution GMRT \HI{} observations,
and the Russian 6-m telescope's Fabry--Perot interferometer data
we address properties of the giant supershell and ionized gas shells related to 
the regions of recent star formation in the galaxy. 
We also present a subsample of the most luminous stars with the assumed record-low metallicities.

\section{Observational data and reduction}
\label{sec:obs}

DDO\,68 was observed with the \textit{HST} using the Advanced Camera for
Surveys (ACS; GO 11578, PI A.\ Aloisi).
Deep images were obtained with the broad-band filters \textit{F606W} (7644\,s)
and \textit{F814W} (7644\,s).
Also, the galaxy was observed with the narrow-band filter \textit{F658N}
centred at \Halpha{} (2388\,s).
The ACS image in \textit{F606W} filter is shown in Fig.\,\ref{fig:ima}.

\begin{figure}
\centerline{
\includegraphics[width=0.35\textwidth,clip]{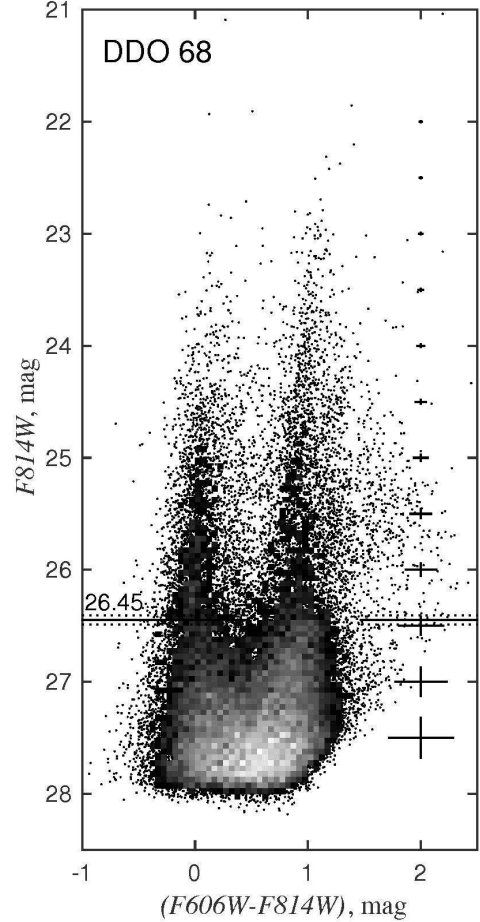} 
}
\caption{
The colour--magnitude diagram of 35786 stars in DDO\,68
in the \textit{HST}/ACS flying system.
The illustration combines a simple plot of individual stars as dots
and a Hess diagram for the most dense regions in grey-scale,
where the denser bin is lighter.
}
\label{fig:cmd}
\end{figure}

We use the ACS module of the
\textsc{DOLPHOT}\footnote{\url{http://americano.dolphinsim.com/dolphot}}
software package by \citet{DolPhot} to perform the photometry of the resolved stars
as well as to run artificial star tests to characterize the completeness
and uncertainties in the measurements.
The data quality images were used to mask bad pixels.
Only stars with photometry of a good quality were used in the analysis.
We have selected about 36000 stars with a signal-to-noise of at least 5
in both filters and $\vert\textrm{sharp}\vert\le0.3$.
The resulting colour--magnitude diagram (CMD) in
$\textit{F606W}-\textit{F814W}$ versus \textit{F814W} is plotted
in Fig.\,\ref{fig:cmd}.

\section{Analysis}
\label{sec:analysis}

\subsection{Colour--magnitude diagram}

The CMD of DDO\,68 is typical for
a dwarf irregular galaxy with ongoing star formation.
The highly populated upper part of the main sequence (MS), blue supergiants,
has a mean colour index close to zero,
while the red supergiants (RSG) and the young asymptotic giant branch (AGB) stars,
as well as the middle-age AGB stars have a colour index 
$(\textit{F606W}-\textit{F814W}) \ga 0.9$
and an \textit{F814W} magnitude brighter than 26.5.
The deep images allow us to resolve the most densely populated red giant branch (RGB),
which is visible at the bottom of the diagram
[$(\textit{F606W}-\textit{F814W}) \ga 0.5$ and $\textit{F814W}\ga26.5$].
We estimate the distance modulus to be equal to \modulus{}
corresponding to a distance of \distance{}
(see Section~\ref{sec:distance}).

\begin{figure}
\centerline{
\begin{tabular}{p{0.43\textwidth}@{\hspace{1mm}}p{0.07\textwidth}}
\includegraphics[width=0.43\textwidth,clip]{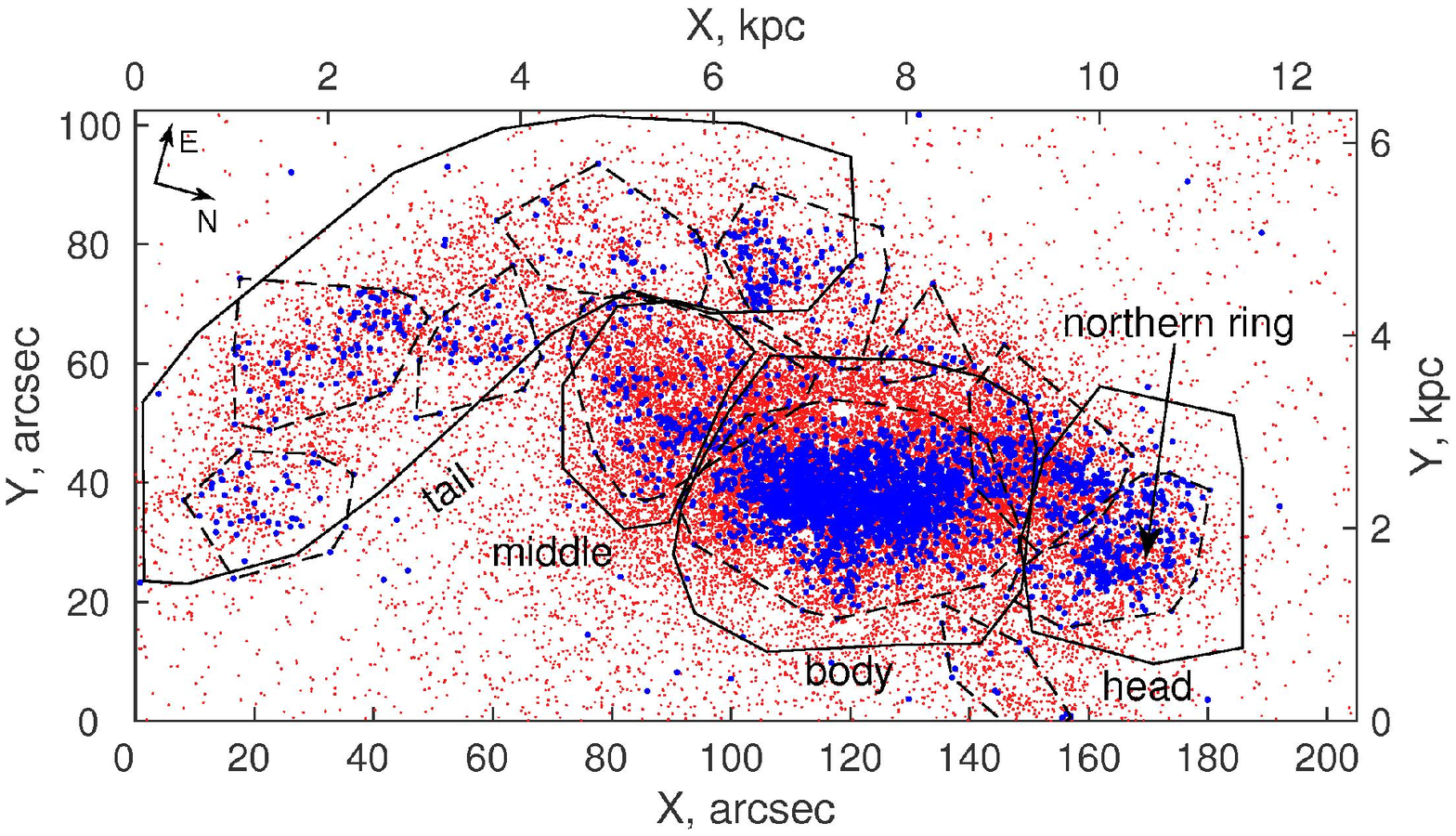} & \raisebox{1.65cm}{ \includegraphics[width=0.05\textwidth,clip]{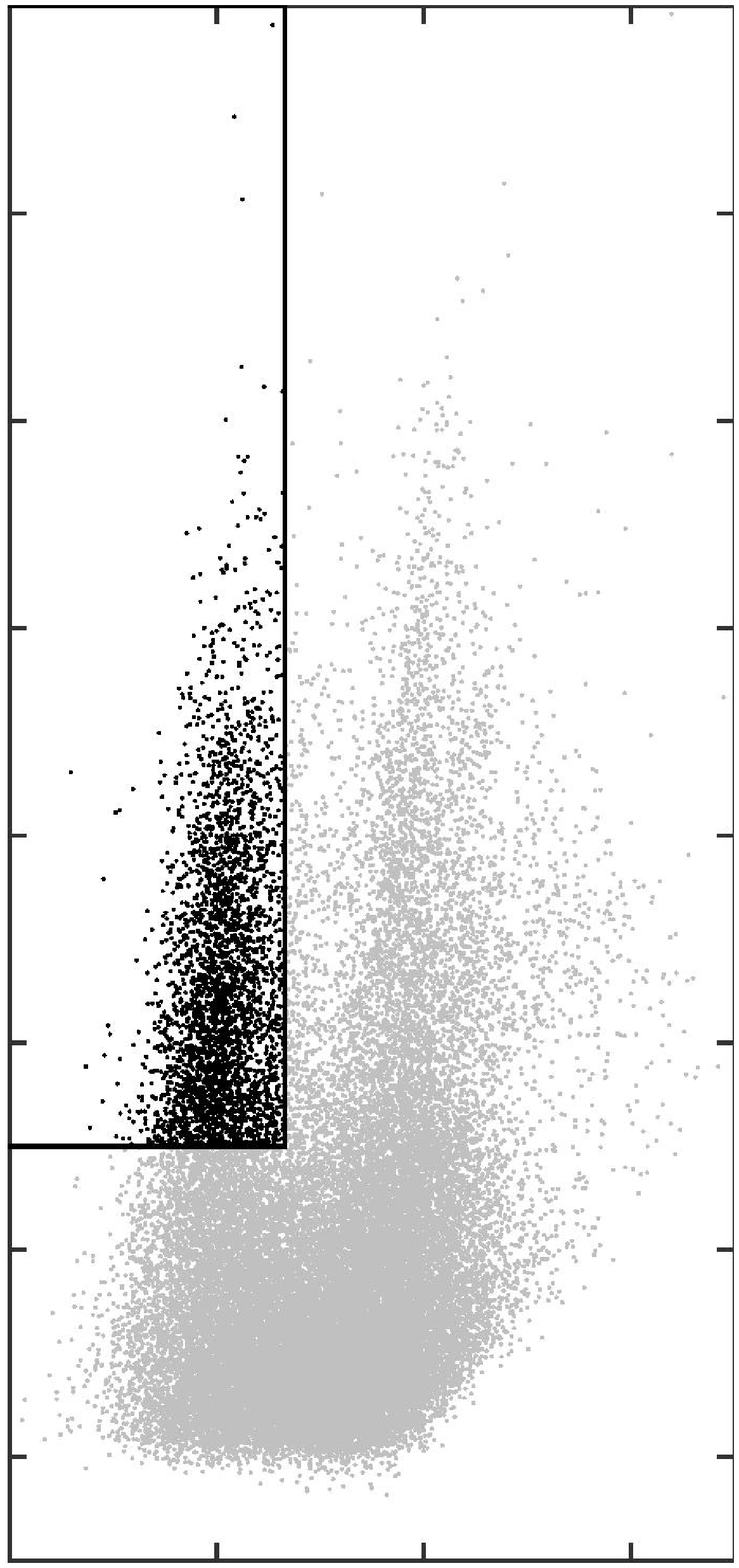} } \\
\end{tabular}
}
\caption{
Distribution of stars in DDO\,68.
Blue points correspond to the bright blue part of the main sequence with
$(\textit{F606W}-\textit{F814W})<0.33$ and $\textit{F814W}<26.5$\,mag,
which is highlighted in the CMD at the right.
Small red dots show the rest of the DDO\,68 stellar populations.
Concentrations of young main-sequence stars are encircled by dashed lines.
The four main regions of DDO\,68 considered in the paper are delineated by
solid lines and marked by their designations.
}
\label{fig:xymap}
\end{figure}

Because DDO\,68 has a complex morphology, for further analysis
we distinguish several parts of the galaxy: `head', `body', `middle', `tail' (see Fig.\,\ref{fig:xymap}).
The `head' (in our notation) includes the `northern ring' and
the `tail' coincides with the `southern tail', named by \citet{PKP2005}.
Deep photometry of resolved stars allows us to study the variation of
the stellar populations in each part in sufficiently high detail.
We analyse the CMD of the entire galaxy as well as of each part separately.
Fig.\,\ref{fig:dens} shows that the distribution of the resolved
stellar populations is very inhomogeneous over the galaxy body. The `head'
is clearly dominated by blue stars. The `body' shows some excess of bright
blue and red supergiants. The `tail' mostly has a considerable excess of
blue but fainter stars ($\textit{F814W} \ga 26$\,mag).
The `middle' part reveals a clear dominance of red giants and AGB stars.
This difference clearly demonstrates a complex evolution of the galaxy
in time and may also favour the hypothesis of DDO\,68 formation
through a merger of two dwarf galaxies.

\subsection{Distance determination}
\label{sec:distance}

The precise distance is a crucial point for the determination of the physical properties
of a galaxy, and, in particular, for the SFH analysis.
\citet{MK1998} measured a distance of 5.9\,Mpc using the brightest blue stars in DDO\,68.
Later, \citet{PT2011} increased it to 9.9\,Mpc
based on the model of the velocity field in the local Universe \citep{TSK2008} .
Recently, \citet{TGL2014} shifted the galaxy significantly farther away to $12.0\pm0.3$\,Mpc
using the TRGB methodology proposed by \citet{LFM1993}.
\citet{DDO68C} received even slightly higher TRGB distance of $12.74\pm0.27$\,Mpc,
but without a specification of the used calibration.
\citet{SAC+2016} found a TRGB modulus of $(m-M)_0=30.41\pm0.12$ ($D=12.08\pm0.67$\,Mpc) 
using the relation between TRGB magnitude and metallicity by \citet{Bellazzini+2004}.
However, they adopt the slightly higher distance modulus $(m-M)_0 = 30.51$, or $D=12.65$\,Mpc, 
as required by the fit of the observations by the synthetic CMD.
The last three TRGB measurements are based on the same deep \textit{HST}/ACS images of DDO\,68
from the \textit{HST} archive (proposal 11578, PI A.\ Aloisi).

We reprocessed these data independently with the \textsc{trgbtool} program,
which uses a maximum-likelihood algorithm for TRGB determination
from the luminosity function of the resolved stellar populations \citep{TRGB1}.
We derived the tip value of \Mtrgb{} in the ACS instrumental system.
The corresponding colour at the position of the TRGB is \Ctrgb{}.
Taking into account the galactic colour excess $ E(B-V) = 0.0183$ from \citet{DustMap},
we derived the true distance modulus of DDO\,68, \modulusLong{}.
It corresponds to the linear distance of \distance{}.
The resulting uncertainty of the distance determination consists of
the very small internal TRGB measurement error of 0.007\,mag,
the calibration error of 0.02\,mag, the foreground extinction inaccuracy
in the direction of DDO\,68 of 0.006\,mag,
and the aperture correction error of 0.05\,mag.
Our distance estimation is in excellent agreement with the measurement of \citet{DDO68C} 
and within the errors of the data of \citet{SAC+2016}.

\subsection{Stellar-population distribution}

\begin{figure}
{\centering
\begin{tabular}{p{0.4\textwidth}@{}p{0.07\textwidth}}
  \includegraphics[width=0.4\textwidth,clip]{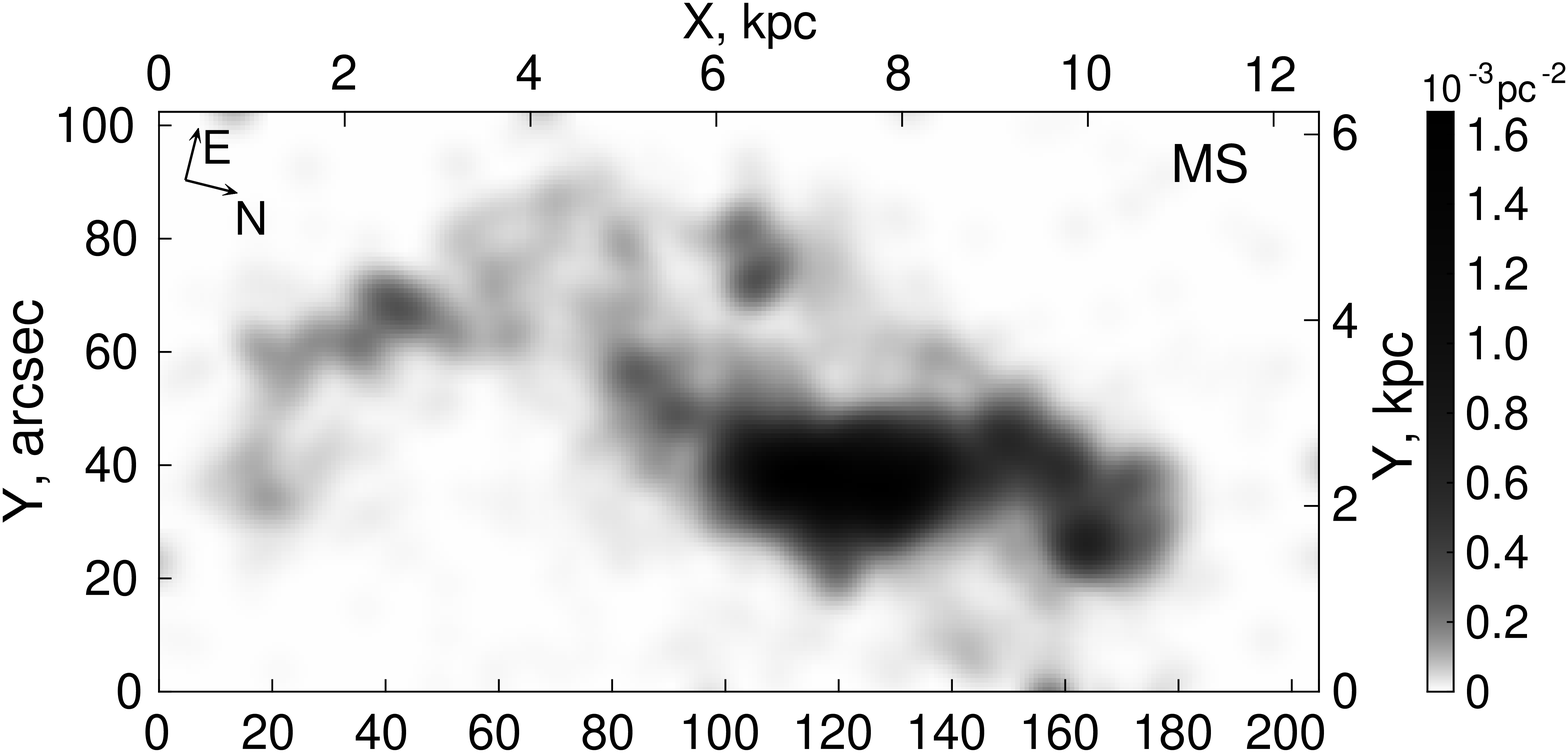}  & \raisebox{0.75cm}{ \includegraphics[width=0.05\textwidth,clip]{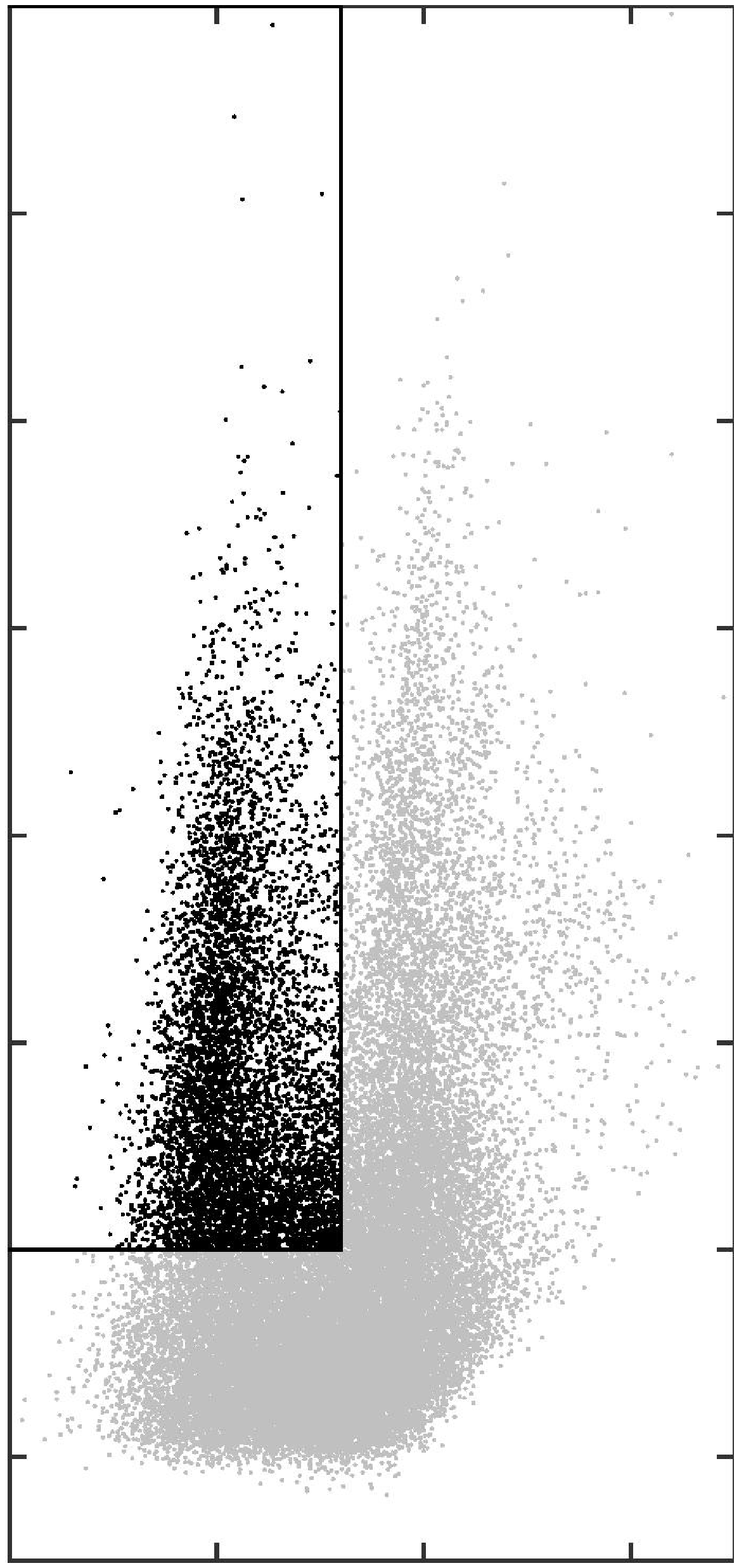}  } \\
  \includegraphics[width=0.4\textwidth,clip]{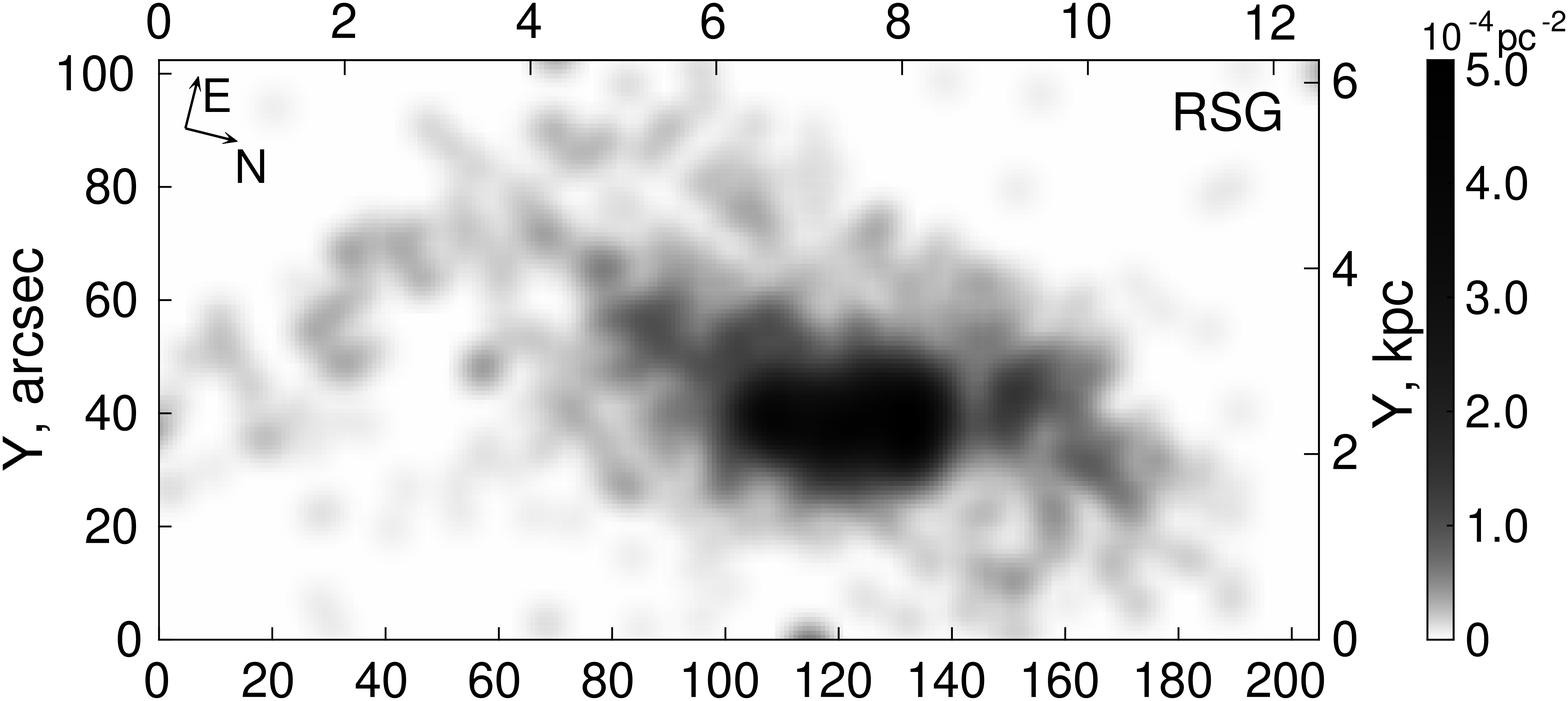} & \raisebox{0.75cm}{ \includegraphics[width=0.05\textwidth,clip]{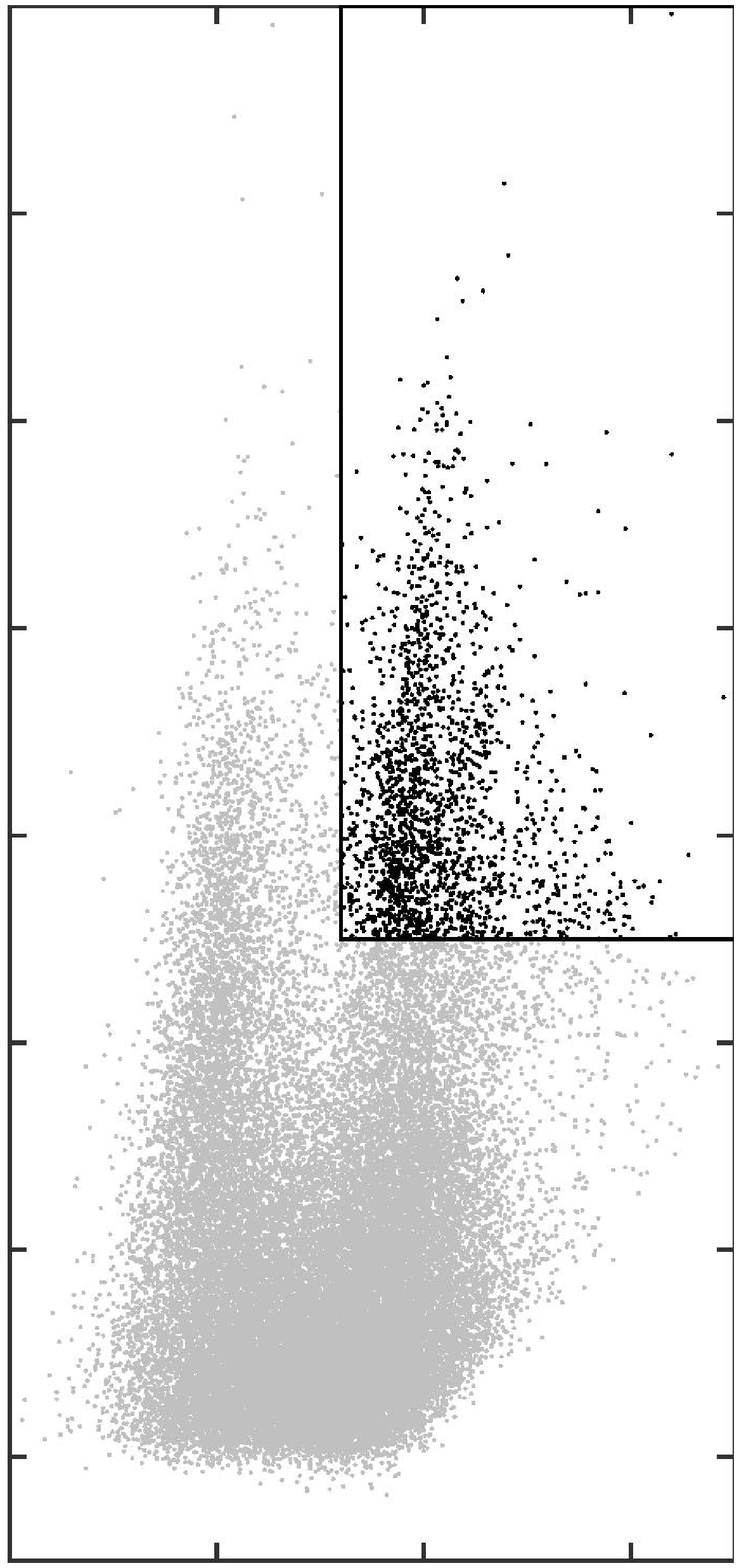} } \\
  \includegraphics[width=0.4\textwidth,clip]{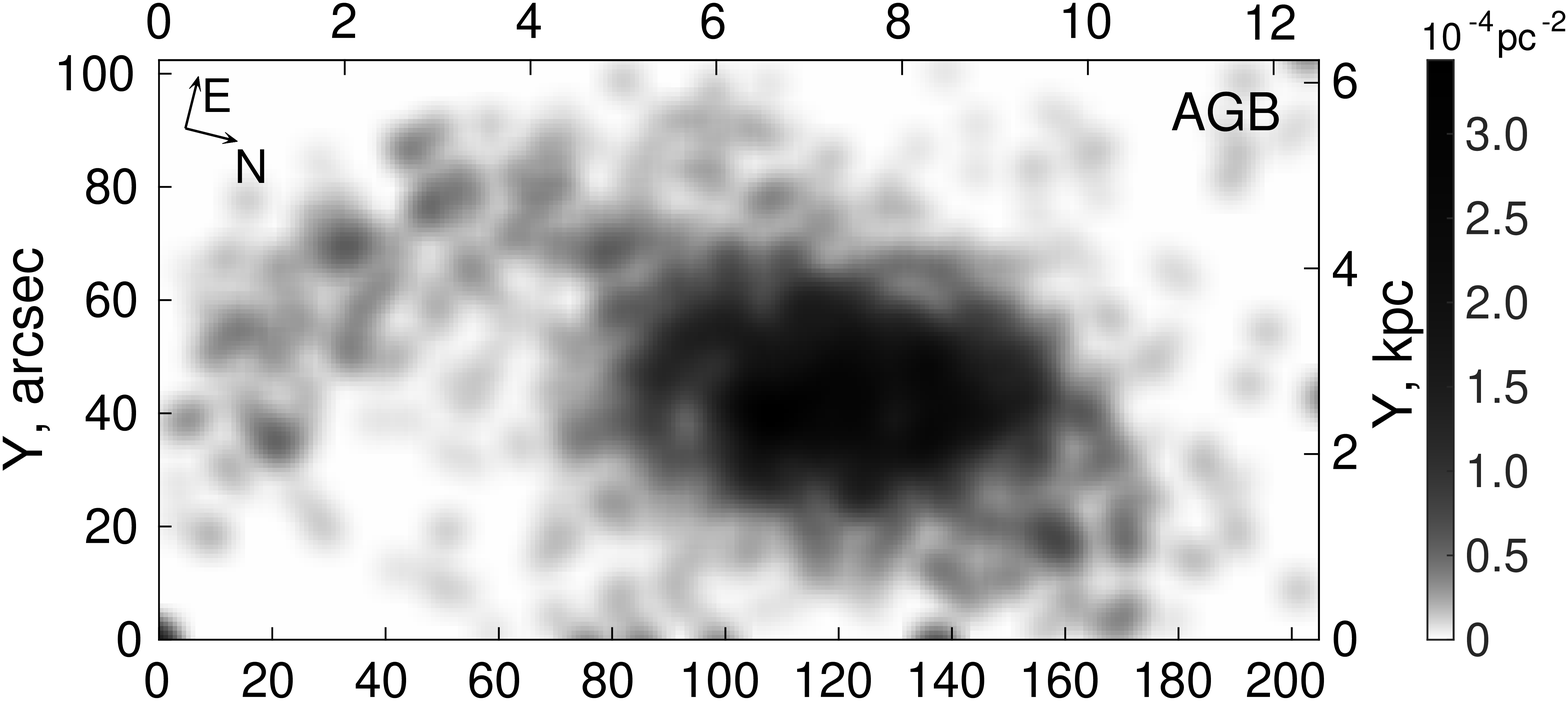} & \raisebox{0.75cm}{ \includegraphics[width=0.05\textwidth,clip]{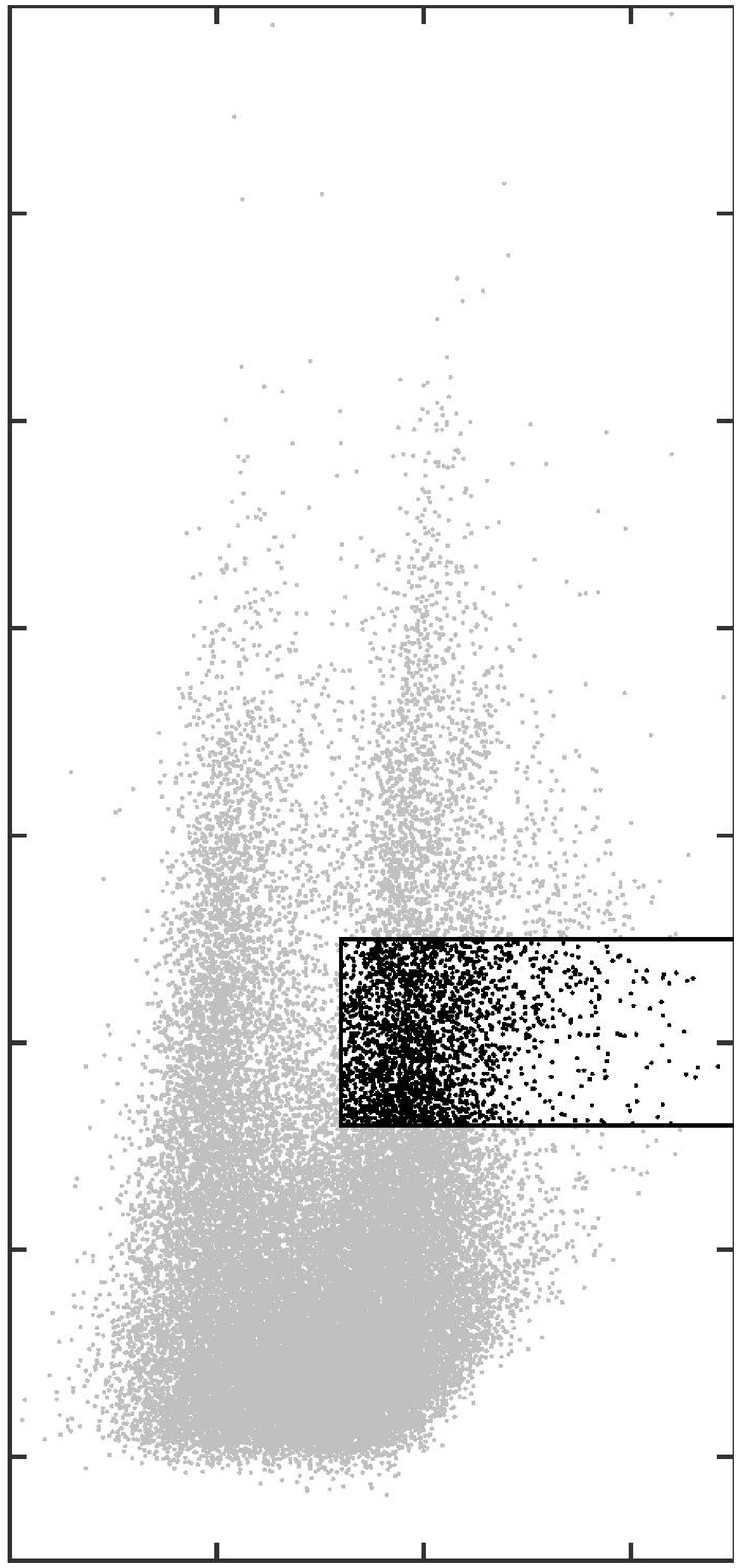} } \\
  \includegraphics[width=0.4\textwidth,clip]{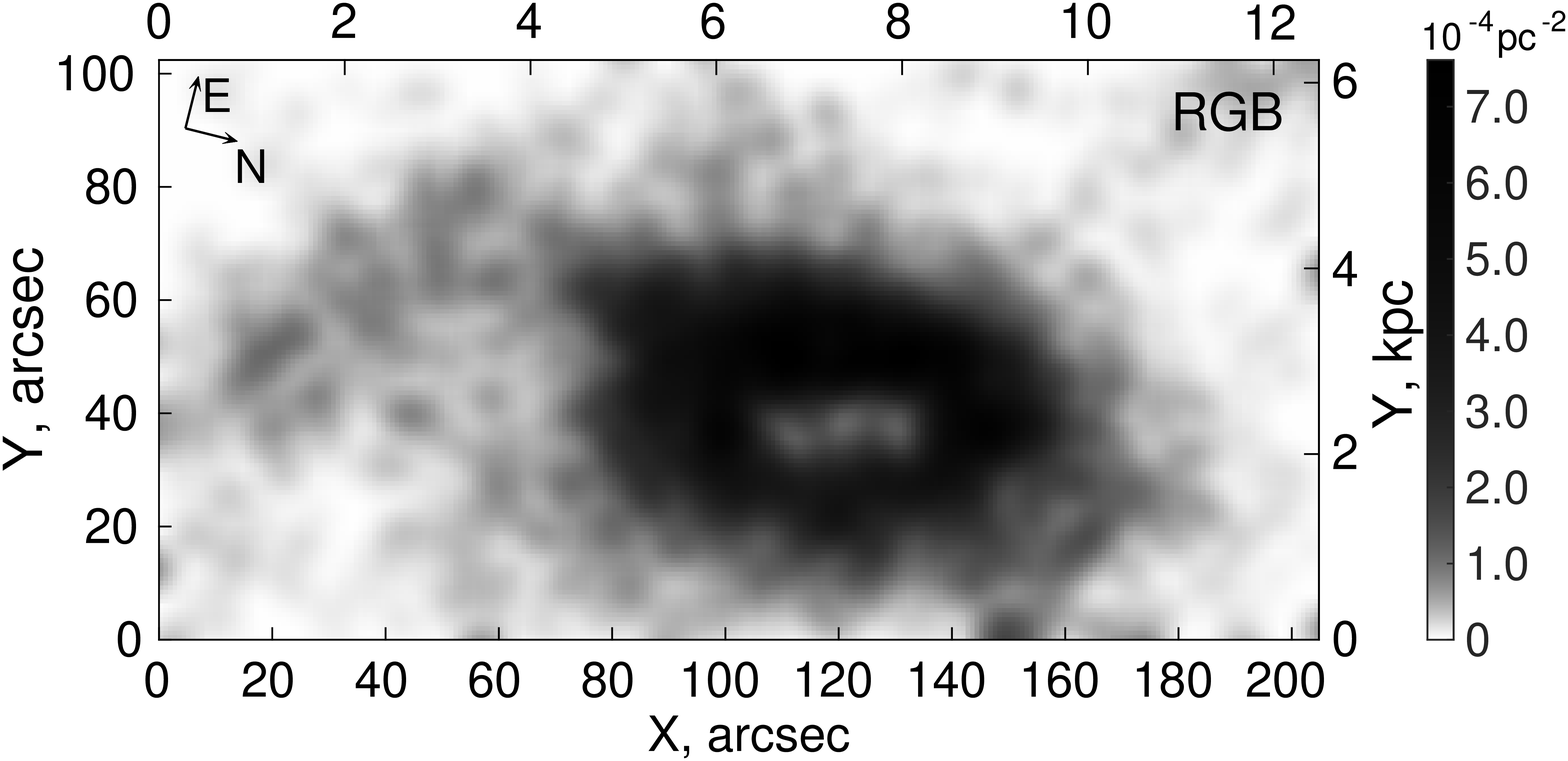} & \raisebox{1.02cm}{ \includegraphics[width=0.05\textwidth,clip]{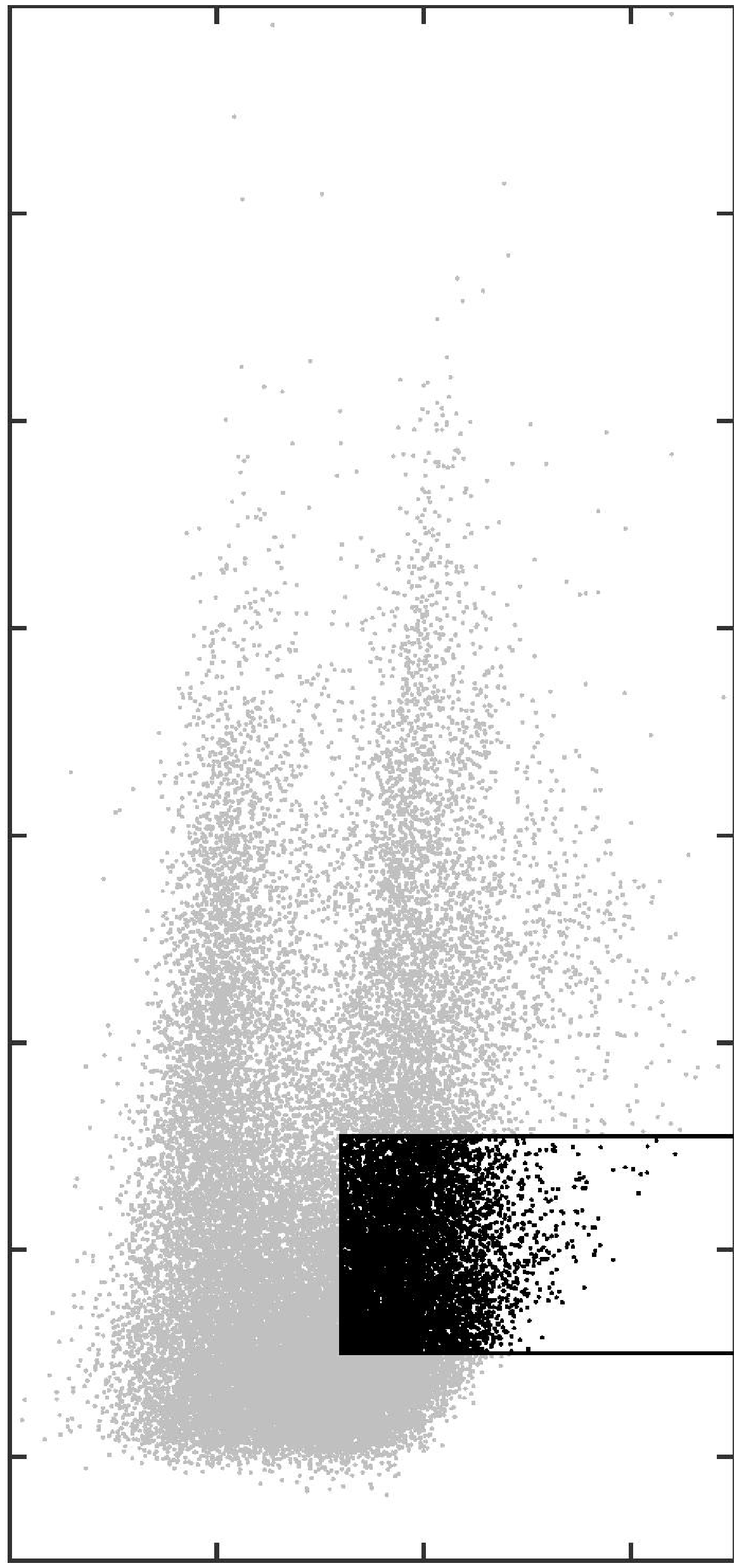} } \\
\end{tabular}
}
\caption{
The maps of density distribution of the four different stellar populations.
The top panel shows MS stars;
the second top panel -- RSG;
the third one from the top -- AGB stars;
and the bottom panel -- RGB population.
The populations selected for the density maps are highlighted
in the respective CMDs at the right.
}
\label{fig:dens}
\end{figure}

In Fig.\,\ref{fig:dens}, we present the density maps of four types of stellar
populations in DDO\,68.
North is approximately to the right, East is up.
The top panel shows the spatial distribution of the MS stars: 
$\textit{F814W}<27.0$ and $(\textit{F606W}-\textit{F814W})<0.6$.
The second top panel depicts the RSG: 
$\textit{F814W}<25.5$ and $(\textit{F606W}-\textit{F814W})>0.6$.
The next panel presents the AGB: 
$25.5<\textit{F814W}<26.4$ and $(\textit{F606W}-\textit{F814W})>0.6$.
The bottom panel shows the distribution of the RGB: 
$26.45<\textit{F814W}<27.5$ and $(\textit{F606W}-\textit{F814W})>0.6$.

The difference in the distribution of the stellar populations of
various ages is seen clearly.
The young stars concentrate in several regions of active star formation.
The most prominent burst coincides with the main body of DDO\,68.
As noted in \citet{PKP2005}, several prominent star formation regions are
well-distinguished in the \textit{HST} \Halpha{} image.
They wind asymmetrically around the bright part of the galaxy
and probably represent the tidal debris of a galaxy minor merger.
This morphology is clearly visible for all stellar populations,
which favours its origin due to a recent interaction.
Moreover, the MS stars show the distortion of the main body of DDO\,68
connecting it with the `northern ring' and the `southern tail' regions (see Fig.\,\ref{fig:dens}).
The distribution of RGB stars shows a `hole' in the centre of DDO\,68 due to the loss of stars in a crowded stellar field.

For the detailed analysis of different stellar populations,
we selected several prominent regions of star formation
using a hierarchical cluster tree
from the sample of the bright blue part of the MS stars
with $(\textit{F606W}-\textit{F814W})<0.33$ and $\textit{F814W}<26.5$\,mag.
These young stars depict the structures in DDO\,68 more clearly.
The result is shown in the Fig.\,\ref{fig:xymap}.
The selected regions are delineated by dashed lines.
One can distinguish a conditional `arc' of 7--8 regions
with an overdensity of the blue MS stars.
This `arc' stretches asymmetrically beside the main body, starting from the
`northern ring' along the eastern edge of the `main body' and
continuing to the south as a `tail'.
These structures are clearly seen in the early images of DDO\,68
and were used to classify this object as
the interacting system VV\,542 by \citet{vv77}.
Taking into account the marked regions,
we extracted for further analysis the next four groups of stars.
These groups include all stars (not only young) within the areas enclosed
by solid lines.
The `body' includes the galaxy centre (about 15000 stars).
We divided the arc of starforming regions into
the `head' (about 3500 stars), which coincides with the `northern ring',
and the `tail' (about 5600 stars), which includes most of the clumps.
Also, we consider separately the `middle' region (about 3500 stars)
between the main body and the `tail'.
It is rather well detached from the centre of the galaxy,
but is probably not a simple extension of the `tail'.

\subsection{SFH reconstruction}

We derived the SFH of DDO\,68 using the \textsc{starprobe} package \citep{StarProbe}.
The program develops an approximation to the observed distribution of stars in the CMD
using the positive linear combination of synthetic diagrams formed by simple
stellar populations (SSP: a set of single-age and single-metallicity populations).
The details of our approach and the \textsc{starprobe} software
are described by \citet{StarProbe} and \citet{makarova2010}.

The observed data were binned into two-dimensional histograms (Hess diagrams)
giving the number of stars in cells of the CMDs.
The respective synthetic Hess diagrams were constructed from theoretical stellar
isochrones and the adopted initial mass function (IMF).
Each isochrone describes the magnitudes and colours of the SSP
with the particular age and metallicity as a map of probabilities
to find a star in each cell.
We used the Padova2000 set of theoretical isochrones \citep{Padova2000},
and a \citet{Salpeter1955} IMF.
The distance was adopted from this paper (see above)
and the Galactic extinction is from \citet{DustMap}.
The synthetic diagrams were altered by the same incompleteness
and crowding effects and by photometric systematics identical to
those determined for the observations using artificial stars experiments.
We took into account the unresolved binary stars
(binary fraction) to be 30 per\,cent. 
The mass function of the individual stars and the main components
of binary systems was assumed to be the same.
The mass distribution for the second components was taken to be flat
in the range 0.7--1.0 of the main component mass.
The synthetic diagrams covered
the whole range of ages (from 4\,Myr to 14\,Gyr) and metallicities
(from $Z = 0.0001$ to $0.03$).
Hess diagrams of the SSP were combined together for a given time range
under the assumption of a constant star formation rate (SFR) during this period
(2-Gyr step for the populations older than 2\,Gyr;
500-Myr step from 500\,Myr to 2\,Gyr;
and 100-Myr step for populations younger than 500\,Myr).
The isochrones were interpolated in age to avoid discontinuities,
so that the sampled points in the CMD were separated at most by 0.03 mag.
The fitting procedure allowed us to use the boundaries of star formation episodes as free parameters of minimization.
The metallicity resolution was limited by the underlying published isochrones; they were not interpolated in metallicity.
As a result, we constructed a set of 60 synthetic Hess diagrams for seven predefined metallicities.

Using a stepwise approach, we selected only \textrm{statistically} significant episodes of star formation.
\textrm{
It allowed us to avoid problems with `noise' from insignificant variables and improved the quality of the fit.
}
At each iteration, we found an artificial diagram that correlated most strongly with the observational data,
taking into account the variables already included in the regression.
Using the partial Fisher criterion, a variable was included in the regression equation
if its significance was greater than a pre-selected value.
After that, the algorithm found a possibility to exclude previously selected variables from the resulting regression equation.
We iterated until no more variables could be included in the regression.
Therefore, this approach was purely statistical and did not impose any restrictions
on the metallicity changes during the lifetime of a galaxy.
It allowed us to analyse complex systems with an appearance of metal-poor populations in a merging process.

\begin{figure*}
\centerline{
  \includegraphics[height=0.25\textheight,clip]{sfh.eps}
  \includegraphics[height=0.25\textheight,clip]{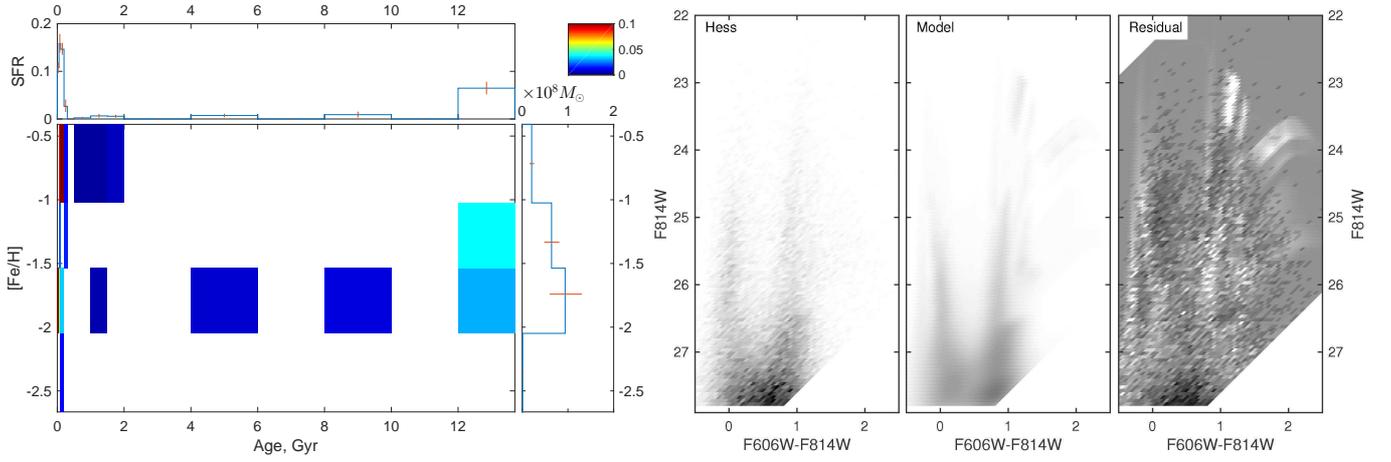}
}
\caption{
Left-hand set:
 the SFH reconstruction of DDO68.
 The measured SFR is coded with a colour (grey-scale) in the central panel
 against the respective age and metallicity.
 The top panel demonstrates the histogram of the SFR versus an age,
 and the left-hand panel shows the total mass fraction versus a metallicity.
 The error bars are $1\sigma$ uncertainties of the maximum-likelihood estimation.
Right-hand set:
 the Hess diagram of the resolved stars of DDO\,68 is shown in the left-hand panel,
 and the best-fitting model Hess diagram is in the right-hand panel.
}
\label{fig:sfh}
\end{figure*}

The best-fitting combination of the synthetic CMDs is a maximum-likelihood solution
taking into account the Poisson noise of the star counts in the cells of the Hess diagram.
The resulting SFH is illustrated in the Fig.\,\ref{fig:sfh}.
The 1$\sigma$ error of each SSP is derived from the analysis of the likelihood function.

According to our measurements, the large burst of star formation in this galaxy
occurred during the period 12--14\,Gyr ago.
This initial burst accounts for 61 per\,cent of the total mass of the formed stars.
Their estimated metallicity is quite low, \FeH{} from $-2$ to $-1$, and the
mean SFR is $7\times10^{-2}$\,\Msunyr{}.
We recognize only very slight traces of star formation
during the periods of about 8--10, 4--6, and 1--2\,Gyr ago.
The detected metallicity of these stars is still low: $\FeH{}\sim[-2:-1.6]$.
The enhanced star formation with a high rate of 0.15 \Msunyr{} occurs in the last 300\,Myr.
The metallicity of these young stars of [$-1$; $-0.5$] is estimated
with significant uncertainties due to their relatively low statistics.
The estimated total stellar mass of DDO\,68 is $M = 1.8\times10^8$ \Msun{}.

Our results are in good agreement with \citet{SAC+2016}, who use a different methodology for SFH reconstruction.
The figs 11--14 from their work show that recent star formation activity increases rapidly about 240--420\,Myr ago
with a peak between $\sim30$ and $\sim50$ Myr ago.
The total stellar mass of young (age $<1$\,Gyr) stars is about $2.8\textrm{--}4.1\times10^{7}$\,\Msun.
This value is in good agreement with our estimation of $3.2\times10^{7}$\,\Msun.

It is worth comparing the SFH in the selected regions of DDO\,68.
We start from the `head' region around the `northern ring'.
There is no sign of ancient star formation.
One can see only insignificant star formation about 8--10\,Gyr ago.
The metallicity of these stars is still within $\FeH = [-1:-2]$.
According to our analysis in this area,
there is no sign of middle-age stars.
The modern episode of star formation is the most prominent in its history.
Although it shows a lower rate than the galaxy as a whole ($1.5\times10^{-2}$\,\Msunyr),
the distribution of the ages and metallicities completely corresponds to it.

Three other regions (the `body', the `middle', and the `tail') show the more or less
oldest substantial episode of star formation in the period of 12--14\,Gyr ago.
The SFR in that epoch is comparable to the ongoing star
formation only in the `middle' region.
In the `body' and the `tail', the SFR is significantly lower than the recent one.
It is interesting to note that the estimated metallicity of the
ancient stars varies from region to region.
It is likely that all stars in the `body' have $\FeH = [-2.0:-1.6]$.
In the `middle' region, the fraction of stars with \FeH{} from $-1.6$ to $-1.0$ is
slightly higher than the population of the lower metallicity stars.
Apparently, the oldest stars in the `tail' have $\FeH=[-1.6:-1.0]$.
Only two of the considered regions, the `middle' and the `body',
show the faint traces of the middle-age stars.
It is worth noting that the ongoing star formation episode (the last 300\,Myr)
is the most active in all considered regions,
with a clear trend towards higher metallicity for most stars.

\begin{figure*}
\centerline{
  \includegraphics[height=0.25\textheight,clip]{head_sfh.eps}
  \includegraphics[height=0.25\textheight,clip]{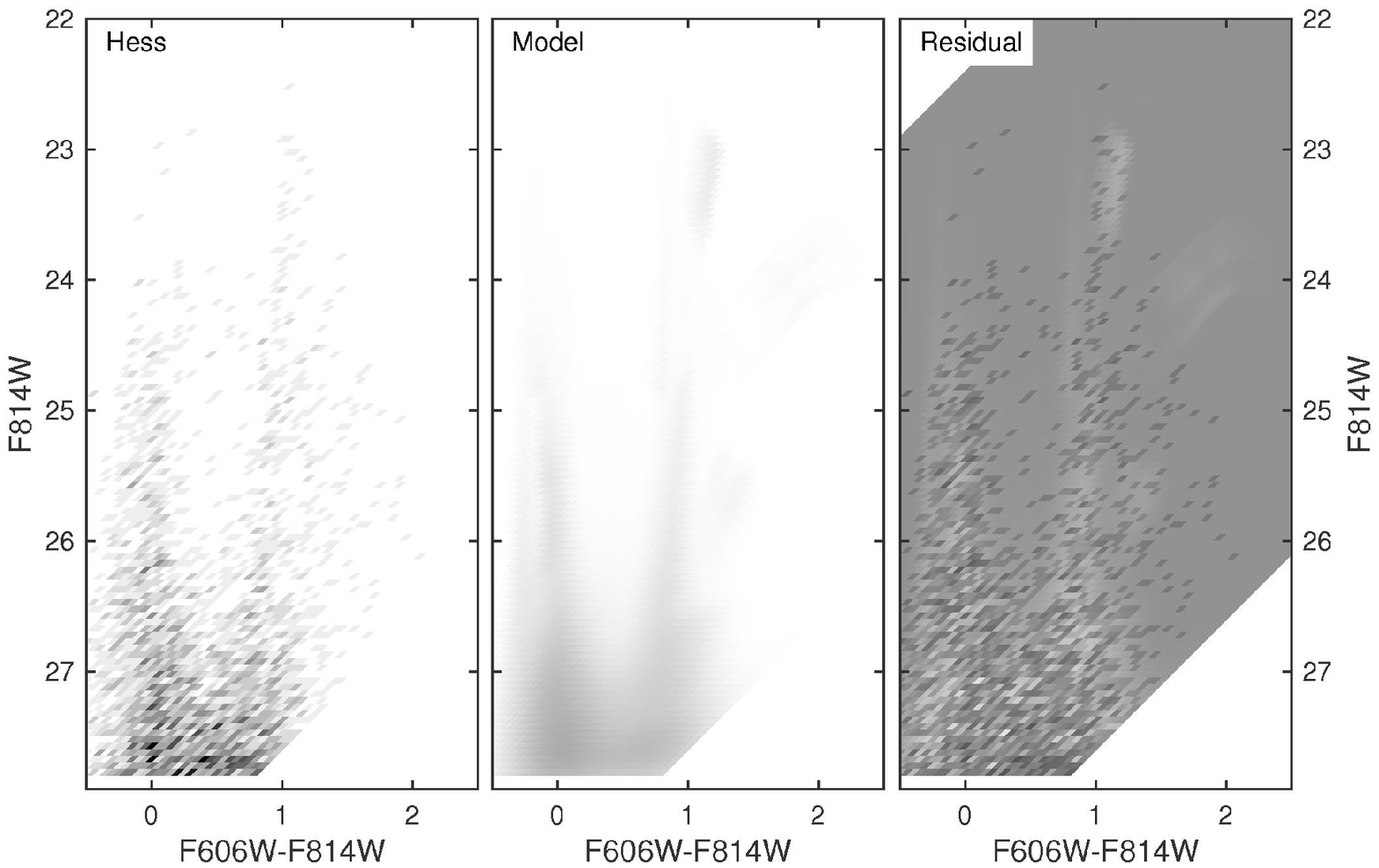}
}
\caption{
SFH reconstruction of `the head' of DDO\,68.
The panels are organized as in Fig.\,\ref{fig:sfh}.
}
\label{fig:sfh-head}
\end{figure*}

\begin{figure*}
\centerline{
  \includegraphics[height=0.25\textheight,clip]{body_sfh.eps}
  \includegraphics[height=0.25\textheight,clip]{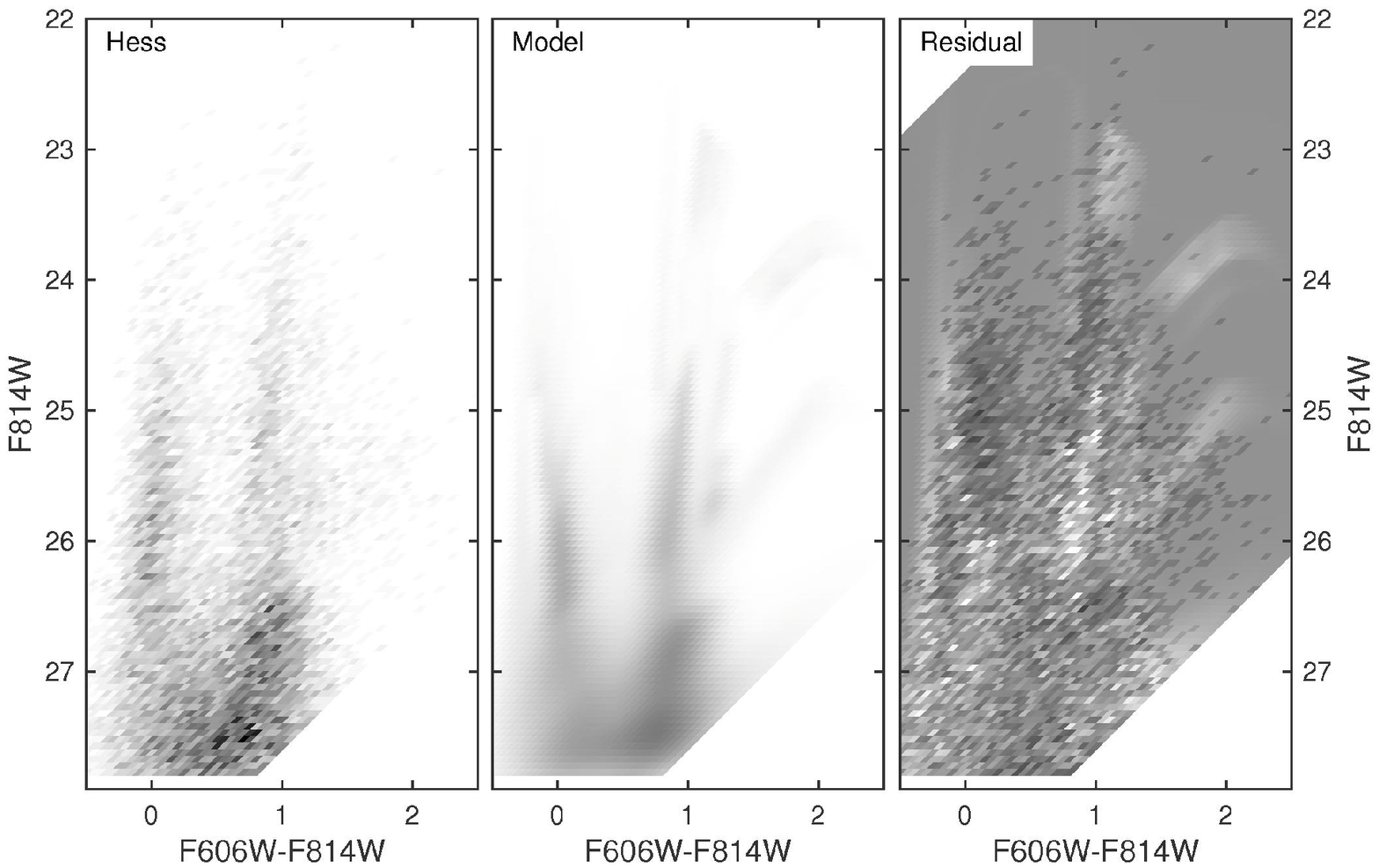}
}
\caption{
SFH reconstruction of `the body' of DDO\,68.
The panels are organized as in Fig.\,\ref{fig:sfh}.
}
\label{fig:sfh-body}
\end{figure*}

\begin{figure*}
\centerline{
  \includegraphics[height=0.25\textheight,clip]{middle_sfh.eps}
  \includegraphics[height=0.25\textheight,clip]{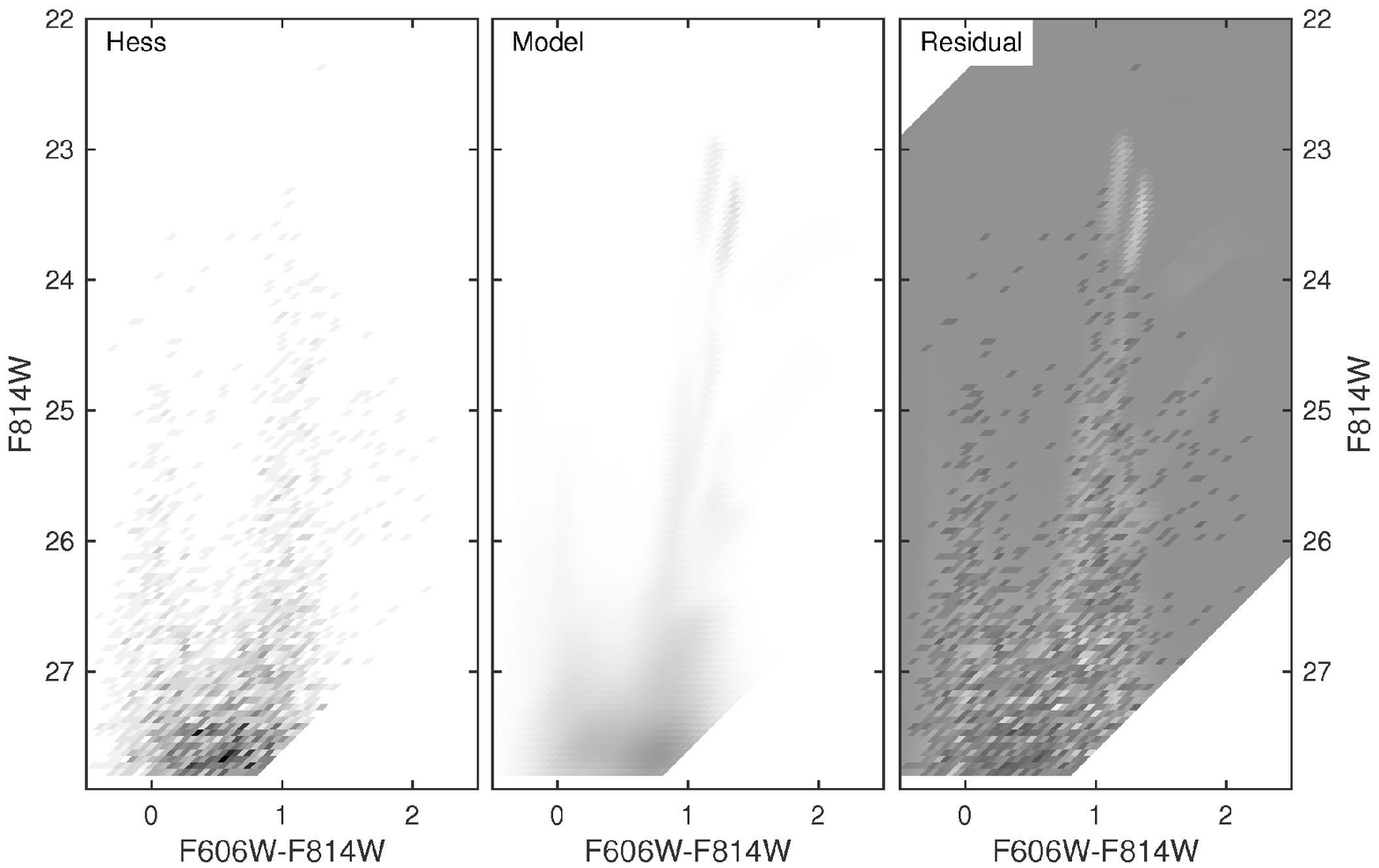}
}
\caption{
SFH reconstruction of `the middle' of DDO\,68.
The panels are organized as in Fig.\,\ref{fig:sfh}.
}
\label{fig:sfh-middle}
\end{figure*}

\begin{figure*}
\centerline{
  \includegraphics[height=0.25\textheight,clip]{tail_sfh.eps}
  \includegraphics[height=0.25\textheight,clip]{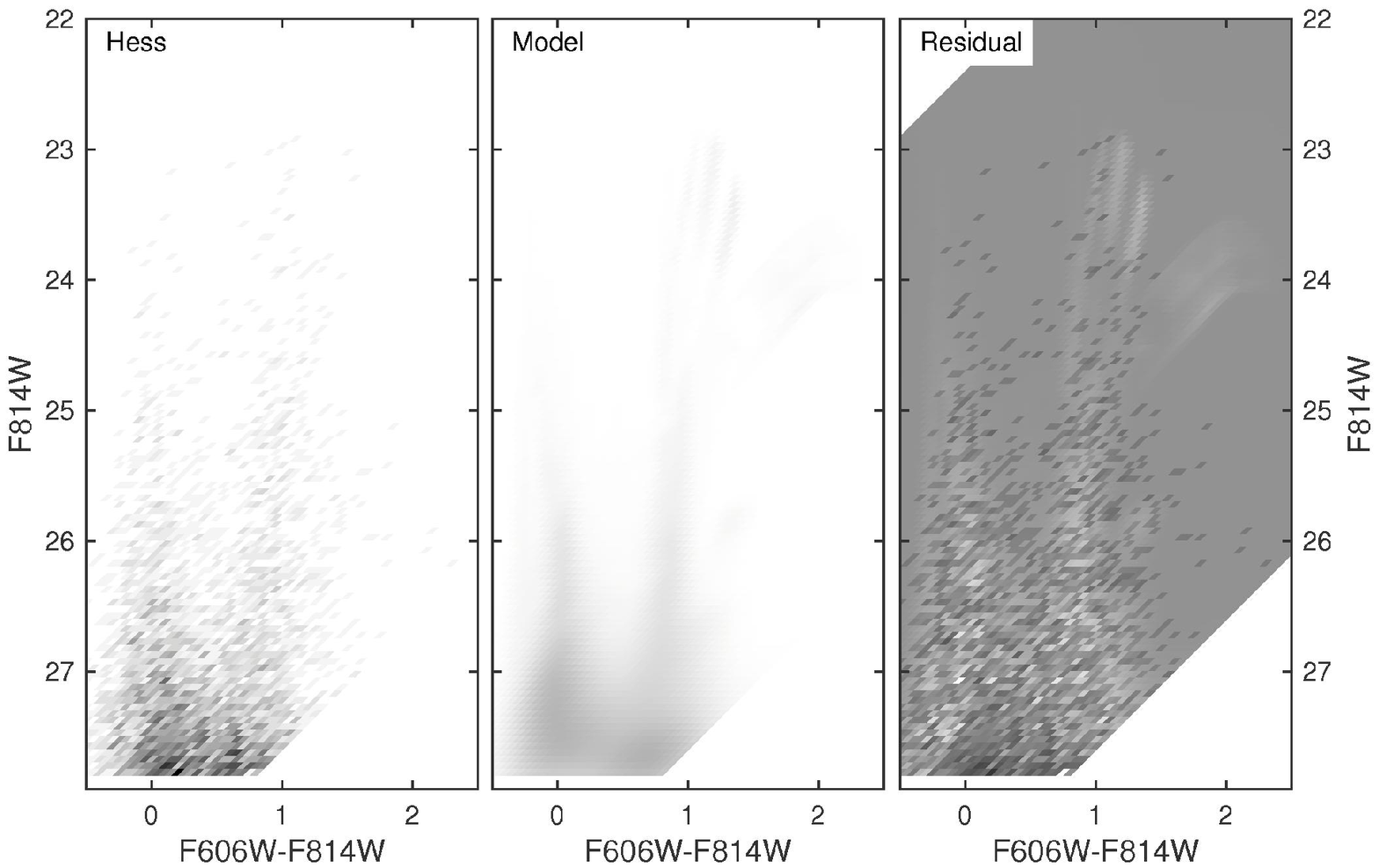}
}
\caption{
SFH reconstruction of `the tail' of DDO\,68.
The panels are organized as in Fig.\,\ref{fig:sfh}.
}
\label{fig:sfh-tail}
\end{figure*}

\section{Discussion}
\label{sec:discus}

\subsection{Old stellar population and its diversity in various parts of DDO\,68}

Our analysis of the CMD of DDO\,68 and the SFH measurements
shows a significant fraction of old giant stars in the `body' region.
The outer regions, the `head' and the `tail', show a higher fraction of
young stars relative to the `body' region.
The main body, hosting a substantial amount of old stars,
extends its population to the outer regions of DDO\,68.
The old stellar populations visible in the `head' region could be explained by
the contribution from the central region.
Thus, we can conclude that the `head' contains only bright and young MS stars.  
In the frame of the minor merger scenario, the `head' region represents the remnant of a destroyed smaller gas-rich component.
With its `young' stellar populations and extremely low $\OH$, this destroyed dwarf is similar to other unusual dwarfs found in the void:
J0926+3343 \citep{PTK2010}, two faint low-surface-brightness dwarfs in the triplet J0723+36 \citep{CP2013}, 
AGC\,198691 \citep{Hirschauer+2016} and J0706+3620 \citep{PPK2016}.

The `tail' region has both young and old stellar populations with an excess of MS stars compared to the 'body'.
The comparison of results for the two regions is evidence
that the stellar population in the `tail' is more evolved than in the `head'.
Therefore, we can conclude that the star formation was induced
during the pre-merger interaction in the `tail' region,
while the star formation burst in the `head' ignited just recently.
The more massive progenitor galaxy has a significant amount
of the old stellar population.
A part of this could be pulled out to the tidal `tail'.
The smaller component of the merger was a gas-rich object with extremely low metallicity, 
as evidenced by the spectra of several young \HII{} regions \citep{PKP2005,IT2007}.
Since the ongoing episode of star formation began about 300\,Myr ago,
we can suggest that the star formation burst marks the epoch of final
coalescence of the merger components.

\subsection{DDO\,68 distance and the problem of large peculiar velocity}

The new DDO\,68 TRGB distance of \distance{} has an important implication.
This distance is 2.85\,Mpc larger than the adopted value by \citet{PT2011}.
The latter distance of 9.9\,Mpc was obtained from the kinematic model
of motion of galaxies in the vicinity of the Local Sheet
due to its repulsion from the huge Local Void \citep{TSK2008}.
The distance estimation takes into account a large negative peculiar velocity
in this sky area of $\Delta V \approx -290$\,\kms.
The derived TRGB distance of DDO\,68 implies an additional negative peculiar velocity
of $-208\pm30$\,\kms{}.
Hence the total negative peculiar velocity for this galaxy is about $-500$\,\kms{}
(for the adopted Hubble constant $H_0=73$\,\kms).
The nature of this large negative peculiar velocity is unclear.

At the projected distances of 0.4--1.5\,Mpc from DDO\,68 there are six galaxies
with very close radial velocities:
J0926+3343 ($\Vh = 536$\,\kms),
KISSB\,23 = J0940+2935 ($\Vh = 505$\,\kms),
UGC\,5209 = J0945+3214 ($\Vh = 538$\,\kms),
UGC\,5272 = J0950+3129 ($\Vh = 520$\,\kms),
J1000+3032 ($\Vh = 484$\,\kms),
and UGC\,5427 = J1004+2921 ($\Vh = 498$\,\kms).
They form a filament with the projected length of $\sim9\degr$,
which corresponds to 1.5--2.0\,Mpc.
The velocity independent distance determination of these galaxies
is crucial to approve this great velocity anomaly
and should help to understand its nature.

One could think that the \citet{TSK2008} kinematic model does not work properly in this region.
However, the recent TRGB distance measurement of UGC\,5209 \citep{KTM2015} is in good agreement with the model.
The TRGB distance of UGC\,5427 presented in the extragalactic distance data base \citep{EDD:CMD} is lower than the model value. 
But a closer look at its CMD reveals probable confusion of RGB stars with the AGB population. 
After the proper correction, the \textit{HST}-based distance of UGC\,5427 also matches the kinematic model. 
The recent detailed theoretical analysis by \citet{Aragon-Calvo+2013} shows
that the velocity field in the void substructures on the scale of 1--2\,Mpc is expected to be very smooth with variations less than 20\,\kms.
However, finding the smooth velocity flow and an abrupt jump in the distance of DDO\,68 
makes the task of independent distance determination for other galaxies in this region very important for understanding the kinematics of filaments in voids.

\section{Summary}
\label{sec:sum}

Summarizing the results and discussion above, we draw the following
conclusions:

\begin{enumerate}
\item
There is a clear dichotomy of the stellar content of DDO\,68 between the
bright main central body and the peripheral regions (including the `head' and the `tail').
Apart from the large fraction of young populations,
the `body' also shows the prominent RGB.
The outer regions display less significant fractions of old stars.
Moreover, these old stars are probably interlopers (at least for `head')
from the outer parts of the main body.
In the frame of this interpretation, the outer parts of DDO\,68 represent
a destroyed, smaller, gas-rich, and
very low-metallicity galaxy in the process of an almost completed merger
with a more massive and more typical dwarf. 

\item
The analysis of the SFH of DDO\,68 shows that 61 per cent
of the total stellar mass was formed during the initial burst
of star formation in the epoch 12--14\,Gyr ago.
There are only weak traces of star formation in the period between 1 and 10\,Gyr.
The modern and ongoing star formation episode started about 300\,Myr ago.
It is characterized by the high mean rate of star formation $\textrm{SFR}=0.15$\,\Msunyr.
The great majority (80--88 per cent) of stars in the system as a whole are metal-poor,
with $Z = Z_{\sun}/50\textrm{--}Z_{\sun}/20$.
Only $\lesssim20$\,per cent of stars have $Z \approx Z_{\sun}/5$.

\item
The new TRGB-based distance of DDO\,68 is $2.8\pm0.4$\,Mpc farther
relative to the previously adopted value from the kinematic model of \citet{TSK2008},
which predicts a large negative peculiar velocity in this region of $\Delta V \approx -290$\,\kms.
This result poses an important question on the origin of the additional
peculiar velocity of $\approx-200$\,\kms{} of DDO\,68,
which, if real, implies a value of the peculiar velocity of
$\approx-500$\,\kms{} in this region.

\end{enumerate}

\section*{Acknowledgements}
This article is based on archival data of
the Space Telescope Science Institute (STScI)
obtained with the NASA/ESA \textit{Hubble Space Telescope}, programme GO--11578.
The STScI is operated by the Association of Universities for Research in
Astronomy, Inc.\ under NASA contract NAS 5--26555.
This work was carried out with the support of the Russian Science Foundation
grant 14--12--00965 (distance and structure of DDO\,68).
LNM acknowledges the support of Research Program OFN-17 of the Division of Physics,
Russian Academy of Sciences (SFH reconstruction).
The authors thank N.\ A.\ Tikhonov, who kindly made available his and his co-authors'
paper prior to publication.


\bibliographystyle{mnras}
\bibliography{ddo68}   

\begin{thebibliography}{}
\makeatletter
\relax
\def\mn@urlcharsother{\let\do\@makeother \do\$\do\&\do\#\do\^\do\_\do\%\do\~}
\def\mn@doi{\begingroup\mn@urlcharsother \@ifnextchar [ {\mn@doi@}
  {\mn@doi@[]}}
\def\mn@doi@[#1]#2{\def\@tempa{#1}\ifx\@tempa\@empty \href
  {http://dx.doi.org/#2} {doi:#2}\else \href {http://dx.doi.org/#2} {#1}\fi
  \endgroup}
\def\mn@eprint#1#2{\mn@eprint@#1:#2::\@nil}
\def\mn@eprint@arXiv#1{\href {http://arxiv.org/abs/#1} {{\tt arXiv:#1}}}
\def\mn@eprint@dblp#1{\href {http://dblp.uni-trier.de/rec/bibtex/#1.xml}
  {dblp:#1}}
\def\mn@eprint@#1:#2:#3:#4\@nil{\def\@tempa {#1}\def\@tempb {#2}\def\@tempc
  {#3}\ifx \@tempc \@empty \let \@tempc \@tempb \let \@tempb \@tempa \fi \ifx
  \@tempb \@empty \def\@tempb {arXiv}\fi \@ifundefined
  {mn@eprint@\@tempb}{\@tempb:\@tempc}{\expandafter \expandafter \csname
  mn@eprint@\@tempb\endcsname \expandafter{\@tempc}}}

\bibitem[\protect\citeauthoryear{{Annibali} et~al.,}{{Annibali}
  et~al.}{2016}]{Annibali+2016}
{Annibali} F.,  et~al., 2016, \mn@doi [\apjl] {10.3847/2041-8205/826/2/L27},
  \href {http://cdsads.u-strasbg.fr/abs/2016ApJ...826L..27A} {826, L27}

\bibitem[\protect\citeauthoryear{{Aragon-Calvo} \& {Szalay}}{{Aragon-Calvo} \&
  {Szalay}}{2013}]{Aragon-Calvo+2013}
{Aragon-Calvo} M.~A.,  {Szalay} A.~S.,  2013, \mn@doi [\mnras]
  {10.1093/mnras/sts281}, \href
  {http://cdsads.u-strasbg.fr/abs/2013MNRAS.428.3409A} {428, 3409}

\bibitem[\protect\citeauthoryear{{Bellazzini}, {Ferraro}, {Sollima}, {Pancino}
  \& {Origlia}}{{Bellazzini} et~al.}{2004}]{Bellazzini+2004}
{Bellazzini} M.,  {Ferraro} F.~R.,  {Sollima} A.,  {Pancino} E.,   {Origlia}
  L.,  2004, \mn@doi [\aap] {10.1051/0004-6361:20035910}, \href
  {http://cdsads.u-strasbg.fr/abs/2004A%26A...424..199B} {424, 199}

\bibitem[\protect\citeauthoryear{{Berg} et~al.,}{{Berg}
  et~al.}{2012}]{Berg+2012}
{Berg} D.~A.,  et~al., 2012, \mn@doi [\apj] {10.1088/0004-637X/754/2/98}, \href
  {http://adsabs.harvard.edu/abs/2012ApJ...754...98B} {754, 98}

\bibitem[\protect\citeauthoryear{{Cannon} et~al.,}{{Cannon}
  et~al.}{2014}]{DDO68C}
{Cannon} J.~M.,  et~al., 2014, \mn@doi [\apjl] {10.1088/2041-8205/787/1/L1},
  \href {http://cdsads.u-strasbg.fr/abs/2014ApJ...787L...1C} {787, L1}

\bibitem[\protect\citeauthoryear{{Chengalur} \& {Pustilnik}}{{Chengalur} \&
  {Pustilnik}}{2013}]{CP2013}
{Chengalur} J.~N.,  {Pustilnik} S.~A.,  2013, \mn@doi [\mnras]
  {10.1093/mnras/sts138}, \href
  {http://cdsads.u-strasbg.fr/abs/2013MNRAS.428.1579C} {428, 1579}

\bibitem[\protect\citeauthoryear{{Chengalur}, {Pustilnik}  \&
  {Egorova}}{{Chengalur} et~al.}{2016}]{CPE2016}
{Chengalur} J.~N.,  {Pustilnik} S.~A.,   {Egorova} E.~S.,  2016, preprint,
  \href {http://cdsads.u-strasbg.fr/abs/2016arXiv161101271C} {} (\mn@eprint
  {arXiv} {1611.01271})

\bibitem[\protect\citeauthoryear{{Contreras Ramos} et~al.,}{{Contreras Ramos}
  et~al.}{2011}]{Contreras+2011}
{Contreras Ramos} R.,  et~al., 2011, \mn@doi [\apj]
  {10.1088/0004-637X/739/2/74}, \href
  {http://adsabs.harvard.edu/abs/2011ApJ...739...74C} {739, 74}

\bibitem[\protect\citeauthoryear{{Dolphin}}{{Dolphin}}{2002}]{DolPhot}
{Dolphin} A.~E.,  2002, \mn@doi [\mnras] {10.1046/j.1365-8711.2002.05271.x},
  \href {http://cdsads.u-strasbg.fr/abs/2002MNRAS.332...91D} {332, 91}

\bibitem[\protect\citeauthoryear{{Ekta,}, {Chengalur}  \& {Pustilnik}}{{Ekta,}
  et~al.}{2008}]{ECP2008}
{Ekta,} {Chengalur} J.~N.,   {Pustilnik} S.~A.,  2008, \mn@doi [\mnras]
  {10.1111/j.1365-2966.2008.13928.x}, \href
  {http://cdsads.u-strasbg.fr/abs/2008MNRAS.391..881E} {391, 881}

\bibitem[\protect\citeauthoryear{{Girardi}, {Bressan}, {Bertelli}  \&
  {Chiosi}}{{Girardi} et~al.}{2000}]{Padova2000}
{Girardi} L.,  {Bressan} A.,  {Bertelli} G.,   {Chiosi} C.,  2000, \mn@doi
  [\aaps] {10.1051/aas:2000126}, \href
  {http://cdsads.u-strasbg.fr/abs/2000A%26AS..141..371G} {141, 371}

\bibitem[\protect\citeauthoryear{{Haynes} et~al.,}{{Haynes}
  et~al.}{2011}]{ALFALFA40}
{Haynes} M.~P.,  et~al., 2011, \mn@doi [\aj] {10.1088/0004-6256/142/5/170},
  \href {http://adsabs.harvard.edu/abs/2011AJ....142..170H} {142, 170}

\bibitem[\protect\citeauthoryear{{Hirschauer} et~al.,}{{Hirschauer}
  et~al.}{2016}]{Hirschauer+2016}
{Hirschauer} A.~S.,  et~al., 2016, \mn@doi [\apj]
  {10.3847/0004-637X/822/2/108}, \href
  {http://adsabs.harvard.edu/abs/2016ApJ...822..108H} {822, 108}

\bibitem[\protect\citeauthoryear{{Izotov} \& {Thuan}}{{Izotov} \&
  {Thuan}}{1998}]{Izotov+1998}
{Izotov} Y.~I.,  {Thuan} T.~X.,  1998, \mn@doi [\apj] {10.1086/305440}, \href
  {http://adsabs.harvard.edu/abs/1998ApJ...497..227I} {497, 227}

\bibitem[\protect\citeauthoryear{{Izotov} \& {Thuan}}{{Izotov} \&
  {Thuan}}{2004}]{IT2004}
{Izotov} Y.~I.,  {Thuan} T.~X.,  2004, \mn@doi [\apj] {10.1086/424990}, \href
  {http://adsabs.harvard.edu/abs/2004ApJ...616..768I} {616, 768}

\bibitem[\protect\citeauthoryear{{Izotov} \& {Thuan}}{{Izotov} \&
  {Thuan}}{2007}]{IT2007}
{Izotov} Y.~I.,  {Thuan} T.~X.,  2007, \mn@doi [\apj] {10.1086/519922}, \href
  {http://cdsads.u-strasbg.fr/abs/2007ApJ...665.1115I} {665, 1115}

\bibitem[\protect\citeauthoryear{{Izotov} \& {Thuan}}{{Izotov} \&
  {Thuan}}{2009}]{IT2009}
{Izotov} Y.~I.,  {Thuan} T.~X.,  2009, \mn@doi [\apj]
  {10.1088/0004-637X/690/2/1797}, \href
  {http://cdsads.u-strasbg.fr/abs/2009ApJ...690.1797I} {690, 1797}

\bibitem[\protect\citeauthoryear{{Izotov}, {Guseva}, {Fricke}  \&
  {Papaderos}}{{Izotov} et~al.}{2009}]{Izotov+2009}
{Izotov} Y.~I.,  {Guseva} N.~G.,  {Fricke} K.~J.,   {Papaderos} P.,  2009,
  \mn@doi [\aap] {10.1051/0004-6361/200911965}, \href
  {http://adsabs.harvard.edu/abs/2009A%26A...503...61I} {503, 61}

\bibitem[\protect\citeauthoryear{{Izotov}, {Thuan}  \& {Guseva}}{{Izotov}
  et~al.}{2012}]{ITG2012}
{Izotov} Y.~I.,  {Thuan} T.~X.,   {Guseva} N.~G.,  2012, \mn@doi [\aap]
  {10.1051/0004-6361/201219733}, \href
  {http://adsabs.harvard.edu/abs/2012A%26A...546A.122I} {546, A122}

\bibitem[\protect\citeauthoryear{{Jacobs}, {Rizzi}, {Tully}, {Shaya}, {Makarov}
   \& {Makarova}}{{Jacobs} et~al.}{2009}]{EDD:CMD}
{Jacobs} B.~A.,  {Rizzi} L.,  {Tully} R.~B.,  {Shaya} E.~J.,  {Makarov} D.~I.,
   {Makarova} L.,  2009, \mn@doi [\aj] {10.1088/0004-6256/138/2/332}, \href
  {http://cdsads.u-strasbg.fr/abs/2009AJ....138..332J} {138, 332}

\bibitem[\protect\citeauthoryear{{Karachentsev}, {Tully}, {Makarova}, {Makarov}
   \& {Rizzi}}{{Karachentsev} et~al.}{2015}]{KTM2015}
{Karachentsev} I.~D.,  {Tully} R.~B.,  {Makarova} L.~N.,  {Makarov} D.~I.,
  {Rizzi} L.,  2015, \mn@doi [\apj] {10.1088/0004-637X/805/2/144}, \href
  {http://cdsads.u-strasbg.fr/abs/2015ApJ...805..144K} {805, 144}

\bibitem[\protect\citeauthoryear{{Lee}, {Freedman}  \& {Madore}}{{Lee}
  et~al.}{1993}]{LFM1993}
{Lee} M.~G.,  {Freedman} W.~L.,   {Madore} B.~F.,  1993, \mn@doi [\apj]
  {10.1086/173334}, \href {http://cdsads.u-strasbg.fr/abs/1993ApJ...417..553L}
  {417, 553}

\bibitem[\protect\citeauthoryear{{Makarov} \& {Makarova}}{{Makarov} \&
  {Makarova}}{2004}]{StarProbe}
{Makarov} D.~I.,  {Makarova} L.~N.,  2004, \mn@doi [Astrophysics]
  {10.1023/B:ASYS.0000031838.50078.1a}, \href
  {http://cdsads.u-strasbg.fr/abs/2004Ap.....47..229M} {47, 229}

\bibitem[\protect\citeauthoryear{{Makarov}, {Makarova}, {Rizzi}, {Tully},
  {Dolphin}, {Sakai}  \& {Shaya}}{{Makarov} et~al.}{2006}]{TRGB1}
{Makarov} D.,  {Makarova} L.,  {Rizzi} L.,  {Tully} R.~B.,  {Dolphin} A.~E.,
  {Sakai} S.,   {Shaya} E.~J.,  2006, \mn@doi [\aj] {10.1086/508925}, \href
  {http://cdsads.u-strasbg.fr/abs/2006AJ....132.2729M} {132, 2729}

\bibitem[\protect\citeauthoryear{{Makarova} \& {Karachentsev}}{{Makarova} \&
  {Karachentsev}}{1998}]{MK1998}
{Makarova} L.~N.,  {Karachentsev} I.~D.,  1998, \mn@doi [\aaps]
  {10.1051/aas:1998315}, \href
  {http://cdsads.u-strasbg.fr/abs/1998A%26AS..133..181M} {133, 181}

\bibitem[\protect\citeauthoryear{{Makarova}, {Koleva}, {Makarov}  \&
  {Prugniel}}{{Makarova} et~al.}{2010}]{makarova2010}
{Makarova} L.,  {Koleva} M.,  {Makarov} D.,   {Prugniel} P.,  2010, \mn@doi
  [\mnras] {10.1111/j.1365-2966.2010.16746.x}, \href
  {http://cdsads.u-strasbg.fr/abs/2010MNRAS.406.1152M} {406, 1152}

\bibitem[\protect\citeauthoryear{{Pustilnik} \& {Tepliakova}}{{Pustilnik} \&
  {Tepliakova}}{2011}]{PT2011}
{Pustilnik} S.~A.,  {Tepliakova} A.~L.,  2011, \mn@doi [\mnras]
  {10.1111/j.1365-2966.2011.18733.x}, \href
  {http://cdsads.u-strasbg.fr/abs/2011MNRAS.415.1188P} {415, 1188}

\bibitem[\protect\citeauthoryear{{Pustilnik}, {Kniazev}  \&
  {Pramskij}}{{Pustilnik} et~al.}{2005}]{PKP2005}
{Pustilnik} S.~A.,  {Kniazev} A.~Y.,   {Pramskij} A.~G.,  2005, \mn@doi [\aap]
  {10.1051/0004-6361:20053102}, \href
  {http://cdsads.u-strasbg.fr/abs/2005A%26A...443...91P} {443, 91}

\bibitem[\protect\citeauthoryear{{Pustilnik}, {Tepliakova}, {Kniazev}  \&
  {Burenkov}}{{Pustilnik} et~al.}{2008}]{DDO68-LBV}
{Pustilnik} S.~A.,  {Tepliakova} A.~L.,  {Kniazev} A.~Y.,   {Burenkov} A.~N.,
  2008, \mn@doi [\mnras] {10.1111/j.1745-3933.2008.00492.x}, \href
  {http://cdsads.u-strasbg.fr/abs/2008MNRAS.388L..24P} {388, L24}

\bibitem[\protect\citeauthoryear{{Pustilnik}, {Tepliakova}, {Kniazev}, {Martin}
   \& {Burenkov}}{{Pustilnik} et~al.}{2010}]{PTK2010}
{Pustilnik} S.~A.,  {Tepliakova} A.~L.,  {Kniazev} A.~Y.,  {Martin} J.-M.,
  {Burenkov} A.~N.,  2010, \mn@doi [\mnras] {10.1111/j.1365-2966.2009.15637.x},
  \href {http://cdsads.u-strasbg.fr/abs/2010MNRAS.401..333P} {401, 333}

\bibitem[\protect\citeauthoryear{{Pustilnik}, {Makarova}, {Perepelitsyna},
  {Moiseev}  \& {Makarov}}{{Pustilnik} et~al.}{2016a}]{DDO68LBV}
{Pustilnik} S.~A.,  {Makarova} L.~N.,  {Perepelitsyna} Y.~A.,  {Moiseev} A.~V.,
    {Makarov} D.~I.,  2016a, preprint, \href
  {http://cdsads.u-strasbg.fr/abs/2016arXiv161108489P} {} (\mn@eprint {arXiv}
  {1611.08489})

\bibitem[\protect\citeauthoryear{{Pustilnik}, {Perepelitsyna}  \&
  {Kniazev}}{{Pustilnik} et~al.}{2016b}]{PPK2016}
{Pustilnik} S.~A.,  {Perepelitsyna} Y.~A.,   {Kniazev} A.~Y.,  2016b, \mn@doi
  [\mnras] {10.1093/mnras/stw2039}, \href
  {http://cdsads.u-strasbg.fr/abs/2016MNRAS.463..670P} {463, 670}

\bibitem[\protect\citeauthoryear{{Rizzi}, {Tully}, {Makarov}, {Makarova},
  {Dolphin}, {Sakai}  \& {Shaya}}{{Rizzi} et~al.}{2007}]{TRGB2}
{Rizzi} L.,  {Tully} R.~B.,  {Makarov} D.,  {Makarova} L.,  {Dolphin} A.~E.,
  {Sakai} S.,   {Shaya} E.~J.,  2007, \mn@doi [\apj] {10.1086/516566}, \href
  {http://cdsads.u-strasbg.fr/abs/2007ApJ...661..815R} {661, 815}

\bibitem[\protect\citeauthoryear{{Sacchi} et~al.,}{{Sacchi}
  et~al.}{2016}]{SAC+2016}
{Sacchi} E.,  et~al., 2016, \mn@doi [\apj] {10.3847/0004-637X/830/1/3}, \href
  {http://cdsads.u-strasbg.fr/abs/2016ApJ...830....3S} {830, 3}

\bibitem[\protect\citeauthoryear{{Salpeter}}{{Salpeter}}{1955}]{Salpeter1955}
{Salpeter} E.~E.,  1955, \mn@doi [\apj] {10.1086/145971}, \href
  {http://cdsads.u-strasbg.fr/abs/1955ApJ...121..161S} {121, 161}

\bibitem[\protect\citeauthoryear{{Schlegel}, {Finkbeiner}  \&
  {Davis}}{{Schlegel} et~al.}{1998}]{DustMap}
{Schlegel} D.~J.,  {Finkbeiner} D.~P.,   {Davis} M.,  1998, \mn@doi [\apj]
  {10.1086/305772}, \href {http://cdsads.u-strasbg.fr/abs/1998ApJ...500..525S}
  {500, 525}

\bibitem[\protect\citeauthoryear{{Skillman} \& {Kennicutt}}{{Skillman} \&
  {Kennicutt}}{1993}]{Skillman+1993}
{Skillman} E.~D.,  {Kennicutt} Jr. R.~C.,  1993, \mn@doi [\apj]
  {10.1086/172868}, \href {http://adsabs.harvard.edu/abs/1993ApJ...411..655S}
  {411, 655}

\bibitem[\protect\citeauthoryear{{Skillman} et~al.,}{{Skillman}
  et~al.}{2013}]{Skillman+2013}
{Skillman} E.~D.,  et~al., 2013, \mn@doi [\aj] {10.1088/0004-6256/146/1/3},
  \href {http://cdsads.u-strasbg.fr/abs/2013AJ....146....3S} {146, 3}

\bibitem[\protect\citeauthoryear{{Stil} \& {Israel}}{{Stil} \&
  {Israel}}{2002}]{SI2002}
{Stil} J.~M.,  {Israel} F.~P.,  2002, \mn@doi [\aap]
  {10.1051/0004-6361:20020352}, \href
  {http://cdsads.u-strasbg.fr/abs/2002A%26A...389...29S} {389, 29}

\bibitem[\protect\citeauthoryear{{Tikhonov}, {Galazutdinova}  \&
  {Lebedev}}{{Tikhonov} et~al.}{2014}]{TGL2014}
{Tikhonov} N.~A.,  {Galazutdinova} O.~A.,   {Lebedev} V.~S.,  2014, \mn@doi
  [Astronomy Letters] {10.1134/S1063773714010058}, \href
  {http://cdsads.u-strasbg.fr/abs/2014AstL...40....1T} {40, 1}

\bibitem[\protect\citeauthoryear{{Tully}, {Shaya}, {Karachentsev}, {Courtois},
  {Kocevski}, {Rizzi}  \& {Peel}}{{Tully} et~al.}{2008}]{TSK2008}
{Tully} R.~B.,  {Shaya} E.~J.,  {Karachentsev} I.~D.,  {Courtois} H.~M.,
  {Kocevski} D.~D.,  {Rizzi} L.,   {Peel} A.,  2008, \mn@doi [\apj]
  {10.1086/527428}, \href {http://cdsads.u-strasbg.fr/abs/2008ApJ...676..184T}
  {676, 184}

\bibitem[\protect\citeauthoryear{{Vorontsov-Velyaminov}}{{Vorontsov-Velyaminov}}{1977}]{vv77}
{Vorontsov-Velyaminov} B.~A.,  1977, \aaps, \href
  {http://cdsads.u-strasbg.fr/abs/1977A%26AS...28....1V} {28, 1}

\makeatother
\end{thebibliography}

\bsp

\label{lastpage}

\end{document}